\documentclass[aps,prb,
 reprint, 
 tightenlines,
 citeautoscript,%
 showpacs,floatfix,
 superscriptaddress]{revtex4-1}

\usepackage{bm}
\usepackage{amssymb}
\usepackage{graphicx}
\usepackage{amsmath}
\usepackage{textcomp}
\usepackage{multirow}

\begin{document}
\title{Numerical study of the optical nonlinearity of doped and gapped
  graphene: From weak to strong field excitation} 
\author{J. L. Cheng}
\affiliation{Brussels Photonics Team (B-PHOT), Department of Applied
  Physics and Photonics (IR-TONA), Vrije Universiteit Brussel,
  Pleinlaan 2, 1050 Brussel, Belgium}
\affiliation{Department of Physics and Institute for Optical Sciences,
  University of Toronto, 60 St. George Street, Toronto, Ontario,
  Canada M5S 1A7}
\author{N. Vermeulen}
\affiliation{Brussels Photonics Team (B-PHOT), Department of Applied
  Physics and Photonics (IR-TONA), Vrije Universiteit Brussel,
  Pleinlaan 2, 1050 Brussel, Belgium}
\author{J. E. Sipe}%
\affiliation{Department of Physics and Institute for Optical Sciences,
University of Toronto, 60 St. George Street, Toronto, Ontario, Canada
M5S 1A7}
\date{\today}
\begin{abstract}
Numerically solving the semiconductor Bloch equations within a 
phenomenological relaxation time approximation, we extract both the linear and
nonlinear optical conductivities of doped graphene and gapped
graphene under excitation by a laser pulse. We discuss in detail the dependence of second harmonic
generation, third harmonic generation, and the Kerr effects on the doping
level, the gap, and the electric field amplitude. The numerical
results for weak electric fields agree with those calculated from available analytic perturbation formulas. For strong
electric fields when saturation effects are important, all the
effective third order nonlinear response coefficients show a strong field dependence. 
\end{abstract}
\maketitle

\section{introduction\label{sec:intro}}
The optical nonlinearity of graphene has been
predicted \cite{Europhys.Lett._79_27002_2007_Mikhailov,Phys.Rep._535_101_2014_Glazov,NewJ.Phys._16_53014_2014_Cheng}
and demonstrated \cite{Phys.Rev.Lett._105_097401_2010_Hendry} to be very strong, which makes graphene an exciting new
candidate for enhancing nonlinear optical functionalities in optical devices
\cite{Nat.Photon._4_611_2010_Bonaccorso,Nat.Photon._6_554_2012_Gu,IEEE_OSAJournalofLightwaveTechnology_30_427_2011_Yamashita,ACSNano_6_3677_2012_Bao}. To
optimize the performance of these devices, one of the preliminary conditions is to fully understand the dependence of the
optical nonlinearity of graphene on the chemical potential \cite{Nat.Photon._8_585_2014_Horiuchi},
temperature, and the excitation frequency. At present, both experiments and
theories are still at an early stage. Experiments have
investigated parametric frequency conversion \cite{Phys.Rev.Lett._105_097401_2010_Hendry}, third harmonic
generation (THG)
\cite{ACSNano_7_8441_2013_Saeynaetjoki,Phys.Rev.B_87_121406_2013_Kumar,Phys.Rev.X_3_021014_2013_Hong},
Kerr effects and two-photon absorption
\cite{NanoLett._11_2622_2011_Yang,Opt.Lett._37_1856_2012_Zhang,Nat.Photon._6_554_2012_Gu,NanoLett._11_5159_2011_Wu},
second harmonic generation (SHG)
\cite{Appl.Phys.Lett._95_261910_2009_Dean,Phys.Rev.B_82_125411_2010_Dean,Phys.Rev.B_85_121413_2012_Bykov,NanoLett._13_2104_2013_An,Phys.Rev.B_89_115310_2014_An,Appl.Phys.Lett._105_151605_2014_Lin},
and two-color coherent control
\cite{NanoLett._10_1293_2010_Sun,NewJ.Phys._14_105012_2012_Sun,Phys.Rev.B_85_165427_2012_Sun}, 
and extracted some third order susceptibilities of graphene 
which are orders of magnitude higher than that of normal metal and
semiconductor materials. However,  the dependence of the nonlinearity
on chemical potential, temperature, and the excitation frequency 
have not been systematically measured.  Of the theoretical studies
reported, most are still at the level of single
particle approximations within different approaches, which include 
perturbative treatments based on Fermi's golden rule
\cite{Phys.Rev.B_83_195406_2011_Rioux,Phys.Rev.B_90_115424_2014_Rioux},
the quasiclassical Boltzmann kinetic approach \cite{Europhys.Lett._79_27002_2007_Mikhailov,J.Phys.Condens.Matter_20_384204_Mikhailov,arXiv:1011.4841,Phys.Rep._535_101_2014_Glazov}, and quantum treatments
based on semiconductor Bloch equations (SBE) or equivalent
strategies
\cite{Opt.Lett._36_4569_2011_Zhang, 
  J.Phys.Condens.Matter._24_205802_2012_Jafari,
  Phys.Rev.B_85_115443_2012_Avetissian,  
  J.Nanophoton._6_61702_2012_Avetissian, 
  Phys.Rev.B_88_165411_2013_Avetissian,
  Phys.Rev.B_88_245411_2013_Avetissian,
  NewJ.Phys._16_53014_2014_Cheng,
  Opt.Express_22_15868_2014_Cheng,
  Phys.Rev.B_90_241301_2014_Mikhailov,
  Phys.Rev.B_91_235320_2015_Cheng,
  arXiv:1506.00534}. 
When optical transitions around the Dirac points dominate, analytic
expressions for the third order conductivities can be obtained
perturbatively by employing the linear dispersion approximation
\cite{NewJ.Phys._16_53014_2014_Cheng,
  Opt.Express_22_15868_2014_Cheng,
  Phys.Rev.B_90_241301_2014_Mikhailov,
  Phys.Rev.B_91_235320_2015_Cheng,
  arXiv:1506.00534}. 
The calculations show that third order conductivities depend strongly
on the chemical potential.  

However, there are discrepancies between experimental results and
theoretical predictions. Using the appropriate experimental parameters, the
 susceptibility values obtained by present theories are orders of magnitude smaller than
measured values
\cite{NewJ.Phys._16_53014_2014_Cheng,Phys.Rev.B_91_235320_2015_Cheng}. Possible
reasons for these discrepancies include: 
(1) the linear dispersion approximation may not be adequate for
determining the third order nonlinearities;
(2) a full band structure calculation beyond the two-band
tight-binding model may be required;
(3) the laser intensity used in experiments may be too strong for a
perturbative approach, with saturation effects becoming important; (4)
thermal effects induced by temperature change and gradients may play a
role in the response, and (5) the inclusion of realistic scattering
and many-body effects may be required even for qualitative agreement
with experiment. At a simpler level, different single-particle theories, even based on equivalent
starting equations at the Dirac cone level, have not reached agreement
on the final expressions for third order conductivities
\cite{NewJ.Phys._16_53014_2014_Cheng,Phys.Rev.B_90_241301_2014_Mikhailov,Phys.Rev.B_91_235320_2015_Cheng,arXiv:1506.00534},
due to the complexity in the analytic calculation. In this work, by
numerically solving SBE in gapped graphene and doped graphene, we address some
of these issues by considering the dependence of the optical
response on the chemical potential and band gap: For weak fields, we
investigate whether or not the perturbative treatment in our previous
work \cite{Phys.Rev.B_91_235320_2015_Cheng} is correct and
adequate, while for strong fields the numerical results enable us to
investigate how saturation can affect the nonlinearity.

We organize this paper as follows: in Sec.~\ref{sec:model}, we
present our model for a gapped graphene; in Sec.~\ref{sec:numeric}, we
present our numerical scheme in the calculation; in
Sec.~\ref{sec:result}, we present our results, which include the
comparison to the available perturbative formulas and the effects of
saturation. We conclude and discuss in Sec.~\ref{sec:con}.  

\section{model\label{sec:model}}
\begin{figure}[h]
\centering
\includegraphics[height=6cm]{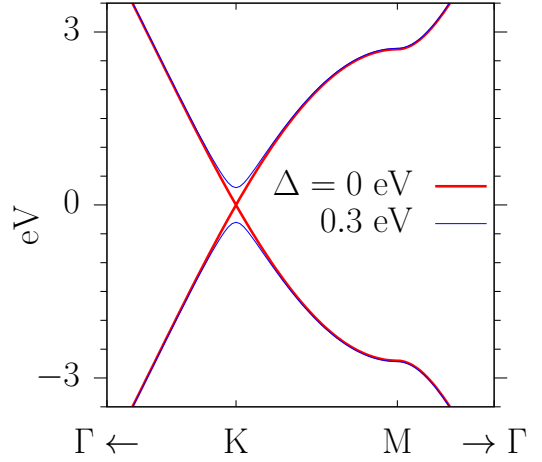}
\caption{(Color online) Band structures of gapped graphene with
  different gap parameter $\Delta=0$ (red thick curves)
  and $0.3$~eV (blue thin curves) at $\Delta=0.3$~eV.} 
\label{fig:band}
\end{figure}

We describe the low energy electronic states by a tight binding model,
employing $p_z$ orbitals $\phi_{\alpha}(\bm r,z)$ with $\alpha=A,B$
for different lattice sites. The band Bloch wave function of the
$s^{th}$ band can be expanded as 
\begin{equation}
  \psi_{s\bm k}(\bm r,z) = \sum_{\alpha}c_{s \bm k}^\alpha
  \Phi_{\alpha\bm k}(\bm r,z)\,,\notag
\end{equation}
where $s$ is the band index, $\bm k=k_x\hat{\bm x} +
k_y\hat{\bm y}$ is the two dimensional wave vector, and the Bloch
state based on site $\alpha$ is
\begin{equation}
\Phi_{\alpha\bm k}(\bm r,z) = (2\pi)^{-1}\sqrt{\Omega}
\sum_{nm}e^{i\bm k\cdot\bm R_{nm}}\phi_{\alpha}(\bm r - \bm R_{nm}-\bm\tau_{\alpha}, z)\,.\notag
\end{equation}
Here $\bm R_{nm}=n\bm a_1 + m\bm a_2$ is the lattice vector, $\Omega$
is the area of one unit cell, $\bm\tau_A=\bm 0$ and
$\bm\tau_B=(\bm a_1+\bm a_2)/3$ are the site positions in one unit
cell, and the primitive lattice vectors $\bm a_i$ are taken as $\bm a_1 = a_0\left(\frac{\sqrt{3}}{2} \hat{\bm x} -
    \frac{1}{2}\hat{\bm y}\right)$ and $\bm a_2 = a_0\left(\frac{\sqrt{3}}{2} \hat{\bm x} +
    \frac{1}{2}\hat{\bm y}\right)$, with the lattice constant $a_0=2.46$~\AA~. In
  our tight binding model, we set the on-site energies as $\Delta$
  for A sites and $-\Delta$ for B sites, the nearest neighbor coupling
  as $\gamma_0=2.7$~eV, and the overlap of the $p_z$ orbitals between
  different sites as zero; the asymmetric on-site energies, resulting
  in a band gap, could be
  induced by a substrate\cite{Nat.Mater._6_916_2007_Zhou}. Then the
  $c_{s \bm k}^\alpha$ satisfy the Schr\"odinger equation 
\begin{equation}
\begin{pmatrix}
  \Delta & \gamma _{0} f_{\bm k} \\ 
\gamma _{0} f_{\bm k}^{\ast } & -\Delta
\end{pmatrix} \begin{pmatrix}
  c_{s\bm k}^{A} \\ c_{s\bm k}^{B}
\end{pmatrix} = \begin{pmatrix}
c_{s\bm k}^{A} \\ 
c_{s\bm k}^{B}%
\end{pmatrix}%
\,.\label{eq:sch}
\end{equation}%
Here $f_{\bm k}=1+e^{-i\bm k\cdot \bm a_{1}}+e^{-i\bm k\cdot \bm a_{2}}$ is
the structure factor. The eigen energies and eigenstates are
\begin{eqnarray}
  \varepsilon_{s\bm k} &=&  s\sqrt{\Delta^2 + (\gamma_0|f_{\bm
      k}|)^2}\,,\quad  s= \pm\,,\notag \notag\\
\begin{pmatrix}
  c_{+\bm k}^{A} \\ c_{+\bm k}^{B}
\end{pmatrix} &=& \frac{1}{\sqrt{2}}\begin{pmatrix} \sqrt{1+{\cal N}_{\bm k}}\\
  \sqrt{1-{\cal N}_{\bm k}}\frac{
  f_{\bm k}^{\ast}}{|f_{\bm k}|}
\end{pmatrix}\,,\notag\\
 \begin{pmatrix}
   c_{-\bm k}^{A} \\ c_{-\bm k}^{B}
\end{pmatrix} &=&\frac{1}{\sqrt{2}} \begin{pmatrix} -\sqrt{1-{\cal N}_{\bm k}}\frac{f_{\bm
      k}}{|f_{\bm k}|} \\ \sqrt{1+{\cal N}_{\bm k}}
\end{pmatrix}\,,\notag
\end{eqnarray}
with ${\cal N}_{\bm k}=  \Delta/\varepsilon_{+\bm
  k}$. 
The band structures for $\Delta=0$ and $0.3$~eV are shown in
Fig.~\ref{fig:band}. For nonzero $\Delta$, the band edges are located at
the Dirac points $\bm K$ and $\bm K^{\prime}$, and the band gap is
$2\Delta$. For $\Delta=0$, gapped graphene reduces to usual
graphene, and the low energy dispersion relation is massless; for nonzero $\Delta$
the low energy dispersion relation is characterized by an effective mass. In the following we call
$\Delta$ the gap parameter.

For later use in the discretization of the derivatives in
Eq.~(\ref{eq:removegauge}), we introduce the matrix elements of
$e^{-i\bm q\cdot\bm r}$ as  
\begin{eqnarray}
  \int d\bm r dz \psi^{\ast}_{s_1\bm k_1}(\bm r,z) e^{-i\bm q\cdot\bm
    r} \psi_{s_2\bm k_2}(\bm r,z) &=& \delta(\bm k_1+\bm q-\bm
  k_2) \notag\\
&&\times U_{s_1\bm k_1;s_2\bm k_1+\bm q}\,,\notag
\end{eqnarray}
Here $U_{s_1\bm k;s_2\bm k+\bm q}$  is calculated by
\begin{equation}
  U_{s_1\bm k;s_2\bm k+\bm q} = \sum_{\alpha_1\alpha_2} \left(c_{s_1\bm
      k}^{\alpha_1}\right)^{\ast} c_{s_2\bm k+\bm q}^{\alpha_2} W_{\alpha_1\bm
    k;\alpha_2\bm k+\bm q}\,.\notag
\end{equation}
with 
\begin{eqnarray}
W_{\alpha_1\bm
    k;\alpha_2\bm k+\bm q}
  &=& \sum_{nm} e^{-i\bm k\cdot\bm R_{nm}} \int_{\Omega}d\bm rdz
  e^{-i\bm q\cdot\bm r} \notag\\
&\times& \phi_{\alpha_1}^{\ast}(\bm r-\bm
  R_{nm}-\bm\tau_{\alpha_1},z) \phi_{\alpha_2}(\bm r-\bm\tau_{\alpha_2},z)\,.\notag
\end{eqnarray}
For small $\bm q$, we approximate $W_{\alpha_1\bm k;\alpha_2\bm
  k+\bm q}\approx e^{-i\bm
  q\cdot\bm\tau_{\alpha_1}}\delta_{\alpha_1\alpha_2}$ which gives
\begin{equation}
U_{s_1\bm k;s_2\bm k+\bm q} = \sum_{\alpha} \left(c_{s_1\bm
    k}^{\alpha}\right)^{\ast} c_{s_2\bm k+\bm q}^{\alpha} e^{-i\bm q\cdot\bm\tau_{\alpha}}\,.
\end{equation}
The Berry connections can be found from $U_{s_1\bm k;s_2\bm k+\bm q}$ by 
\begin{equation}
  \bm \xi_{s_1s_2\bm k} = i\left.\left(\bm\nabla_{\bm q} U_{s_1\bm k;s_2\bm
      k+\bm q}\right) \right|_{\bm q=0}\,,
\end{equation}
and then the velocity matrix elements are given as $\bm v_{ss\bm k} =
\hbar^{-1}\bm\nabla_{\bm k}\varepsilon_{s\bm k}$ and $\bm v_{s\bar{s}\bm
  k}=i\hbar^{-1}(\varepsilon_{s\bm k}-\varepsilon_{\bar{s}\bm
  k})\xi_{s\bar{s}\bm k}$, with $\bar{s}=+(-)$ when $s=-(+)$. After some algebra, we find 
\begin{eqnarray}
  \bm v_{+-\bm k} &=& \left(c_{+\bm k}^{A}\right)^{\ast} c_{-\bm k}^B
  \bm g_{\bm k} + \left(c_{+\bm k}^{B}\right)^{\ast} c_{-\bm k}^A
  \bm g_{\bm k}^{\ast} \notag\\
  &=&\frac{1}{f_{\bm k}}\left\{ i \text{Im}[f_{\bm k}\bm g_{\bm k}] +
    {\cal N}_{\bm k} \text{Re}[f_{\bm k} \bm g_{\bm k}]\right\}\,,
\end{eqnarray}
with $\bm g_{\bm k} = \hbar^{-1}\gamma_0\left[\bm\nabla_{\bm k} f_{\bm k} +i
(\bm\tau_B-\bm \tau_A)f_{\bm k}\right]$. We are interested in optical transitions around the Dirac points $\bm K=(\bm b_1 + 2\bm
b_2)/3$ and $\bm K^{\prime}=(2\bm b_1+\bm b_2)/3$ with the primitive reciprocal lattice vectors $
\bm b_1=\frac{2\pi}{a_0}\left(\frac{1}{\sqrt{3}}\hat{\bm x}-\hat{\bm
    y}\right)$ and $\bm b_2 = \frac{2\pi}{a_0}\left(\frac{1}{\sqrt{3}}\hat{\bm x}+\hat{\bm
    y}\right)$. The usual approximated quantities around the Dirac
points which we used are listed in Table~\ref{tab:app}. 
\begin{table}[t]
  \centering
  \caption{Lowest order approximations for $\bm k$ around the Dirac
    points. Here we use $ v_F=\sqrt{3}a_0\gamma_0/(2\hbar)$, $\bm \kappa= \kappa
    \hat{\bm\kappa}$ with $\hat{\bm \kappa}=\cos\theta \hat{\bm x} + \sin\theta
    \hat{\bm y}$,  $\alpha_{\kappa} = {\hbar v_F
      \kappa}/{\sqrt{\Delta^2 + (\hbar v_F \kappa)^2}}$, and $\beta_{\kappa}={\Delta}/{\sqrt{\Delta^2 + (\hbar v_F \kappa)^2}}$}
  \label{tab:app}
  \begin{tabular}[t]{c|cc|cc}
    \hline\hline
    &\mbox{} & $\bm k = \bm K + \bm \kappa $ & \mbox{} & $\bm k =\bm K^{\prime}+\bm \kappa $
    \\
    \hline
    $\gamma_0 f_{\bm k}$ & \mbox{} & $i \hbar v_F \kappa e^{-i\theta}$
    & \mbox{} &
    $i\hbar v_F \kappa 
    e^{i\theta}$\\
    \hline
    $\varepsilon_{+\bm k}$ &\mbox{} & $\sqrt{(\hbar v_F \kappa)^2+\Delta^2}$,
    &\mbox{} & $\sqrt{(\hbar v_F \kappa)^2+\Delta^2}$\\
    $\bm v_{++\bm k}$ &\mbox{} & $\alpha_{\kappa}v_F\hat{\bm\kappa}$ &\mbox{} &
    $\alpha_{\kappa} v_F\hat{\bm\kappa} $ \\
    \hline
    $\bm g_{\bm k}$ &\mbox{} & $v_F(i\hat{\bm x} + \hat{\bm y})$ &\mbox{} & $v_F
    (i\hat{\bm x} - \hat{\bm y})$\\    
    \hline
    $v^x_{+-\bm k}$ &\mbox{} &
    $v_Fe^{-i\theta}\left(i\beta_{\kappa}\cos\theta-\sin\theta\right)$
    &\mbox{} &  $v_Fe^{i\theta}\left(i\beta_{\kappa}\cos\theta+\sin\theta\right)$ \\
    $v^y_{+-\bm k}$ &\mbox{} &
    $v_Fe^{-i\theta}\left(i\beta_{\kappa}\sin\theta+\cos\theta\right)$
    &\mbox{} &
    $v_Fe^{i\theta}\left(i\beta_{\kappa}\sin\theta-\cos\theta\right)$\\
    \hline\hline
  \end{tabular}
\end{table}

With the application of an external homogeneous electric field $\bm
E(t)$, within the independent particle approximation the time
evolution of the system can be described by SBE
\cite{Phys.Rev.B_91_235320_2015_Cheng} 
\begin{eqnarray}
  i\hbar \frac{\partial \rho _{\bm k}(t)}{\partial t}&=&[\mathcal{E}_{%
\bm k}-e\bm E(t)\cdot \bm\xi _{\bm k},\rho _{\bm k}(t)]-ie\bm
E(t)\cdot \bm\nabla _{\bm k}\rho _{\bm k}(t)\notag\\
&+&i\hbar\left.\frac{\partial
  \rho_{\bm k}(t)}{\partial t}\right|_{\text{scat}}\,.\label{eq:krho}
\end{eqnarray}
Here $\rho_{\bm k}$ is a single particle density matrix, for which the diagonal
term $\rho_{ss\bm k}$ gives the occupation at state $\psi_{s\bm k}$
and the off-diagonal term $\rho_{+-\bm k}$ identifies the interband
polarization between two bands; ${\cal E}_{\bm k}$ is the
energy matrix with elements ${\cal E}_{s_1s_2\bm
  k}=\delta_{s_1s_2}\varepsilon_{s_1\bm k}$; and $e=-|e|$ is the electron
charge. Although $\bm \xi_{\bm k}$ alone is a gauge dependent
quantity, depending on the phases chosen for the Bloch functions, the
combination with the derivative term $\bm\nabla_{\bm k}$ is gauge
independent and can be written as
\begin{eqnarray}
&&[-e\bm E(t)\cdot \bm\xi _{\bm k},\rho _{\bm k}]-ie\bm
E(t)\cdot \bm\nabla _{\bm k}\rho _{\bm k} \notag\\
&=& -ie\bm E(t)\cdot
\left.\bm\nabla_{\bm q}\left(U_{\bm k;\bm k+\bm q}\rho_{\bm k+\bm q}U_{\bm k+\bm
  q;\bm k}\right)\right|_{\bm q=0}\,,\label{eq:removegauge}
\end{eqnarray}
The term $\left.\frac{\partial
  \rho_{\bm k}(t)}{\partial t}\right|_{\text{scat}}$ describes the
relaxation processes. In a phenomenological way we model the intraband (interband)
relaxation process by a parameter $\Gamma_i$ ($\Gamma_e$), and then
\begin{eqnarray}
  \hbar \left.\frac{\partial
  \rho_{ss\bm k}(t)}{\partial t}\right|_{\text{scat}} &=& - \Gamma_i
[\rho_{ss\bm k}(t) - \rho^0_{ss\bm k}]\,,\notag\\
  \hbar \left.\frac{\partial
      \rho_{s\bar{s}\bm k}(t)}{\partial t}\right|_{\text{scat}} &=& - \Gamma_e
  \rho_{s\bar{s}\bm k}(t)\,,\label{eq:kbe}
\end{eqnarray}
where the density matrix at equilibrium state is given by $\rho_{s_1s_2\bm k}^0 = [1+e^{(\varepsilon_{s_1\bm
    k}-\mu)/(k_BT)}]^{-1}\delta_{s_1s_2}$ at temperature $T$ and chemical potential
$\mu$. The current density is calculated as 
\begin{equation}
  \bm J(t) = e\sum_{s_1s_2}\int\frac{d\bm k}{(2\pi)^2}\bm v_{s_2s_1\bm k} \rho_{s_1s_2\bm k}(t)\,.\label{eq:j}
\end{equation}
To focus on the nonlinear response, we separate the linear and
nonlinear contributions to the perturbed density matrix by writing $\rho_{\bm
  k}(t) = \rho_{\bm k}^0 + \rho_{\bm k}^{(1)}(t) + \rho_{\bm
  k}^{(nl)}(t)$ where $\rho_{\bm k}^{(1)}(t)$  is the perturbative
linear contribution of the electric field and determined by  
\begin{eqnarray}
  i\hbar \frac{\partial \rho^{(1)}_{\bm k}(t)}{\partial t}&=&[\mathcal{E}_{%
    \bm k}, \rho^{(1)}_{\bm k}(t)] -e\bm E(t)\cdot \{ [\bm\xi _{\bm k},\rho^{0}_{\bm k}]+i\bm\nabla _{\bm k}\rho^{0}_{\bm k}\}\notag\\
&&-i \begin{pmatrix}
    \Gamma_i \rho_{++\bm k}^{(1)}(t) &     \Gamma_e \rho_{+-\bm
      k}^{(1)}(t) \\
    \Gamma_e \rho_{-+\bm k}^{(1)}(t) &     \Gamma_i \rho_{--\bm
      k}^{(1)}(t) 
  \end{pmatrix}\,,\label{eq:kbelinear}
\end{eqnarray}
while $\rho_{\bm k}^{(nl)}(t)$ includes all higher order contributions
and satisfies the equation
\begin{eqnarray}
  i\hbar \frac{\partial \rho^{(nl)}_{\bm k}(t)}{\partial t}&=&[\mathcal{E}_{%
    \bm k}, \rho^{(nl)}_{\bm k}(t)] -e\bm E(t)\cdot \Big\{ [\bm\xi _{\bm
    k},\rho^{(1)}_{\bm k}(t)+\rho^{(nl)}_{\bm k}(t)]\notag\\
&&+i\bm\nabla _{\bm
    k}[\rho^{(1)}_{\bm k}(t)+\rho^{(nl)}_{\bm k}(t)]\Big\}\notag\\
&&-i \begin{pmatrix}
    \Gamma_i \rho_{++\bm k}^{(nl)}(t) &     \Gamma_e \rho_{+-\bm
      k}^{(nl)}(t) \\
    \Gamma_e \rho_{-+\bm k}^{(nl)}(t) &     \Gamma_i \rho_{--\bm
      k}^{(nl)}(t) 
  \end{pmatrix}\,,\label{eq:kbenonlinear}
\end{eqnarray}
The solution of Eqs.~(\ref{eq:kbelinear}) and (\ref{eq:kbenonlinear})
completely determines the evolution of the single-particle density
matrix, and the current can be written as $J^d(t) = 
J^{(1);d}(t) + J^{(nl);d}(t)$ where $J^{(1);d}(t)$ and $J^{(nl);d}(t)$
are induced by $\rho^{(1)}_{\bm k}(t)$ and $\rho^{(nl)}_{\bm k}(t)$
respectively, and describe the linear and nonlinear response.

\section{Numerical scheme and fitting procedure\label{sec:numeric}}
We consider the response of the current to an applied electric field
pulse with a Gaussian envelope function,
\begin{equation}
\bm E(t) = \hat{\bm x}E_0 e^{-t^2/\Delta_c^2} e^{-i\omega_c t} +
c.c.\,, \label{eq:elf}
\end{equation}
with a duration $\Delta_c$ and a center frequency
$\omega_c$. In the frequency domain, this corresponds to a function
with Gaussian peaks at $\pm\omega_c$
\begin{eqnarray}
\bm E(\omega) &=&\int dt
 e^{i\omega t} \bm E(t) \notag\\
&=&\sqrt{\pi}\Delta_c E_0
 \left[e^{-(\omega-\omega_c)^2\Delta_c^2/4} +
 e^{-(\omega+\omega_c)^2\Delta_c^2/4} \right]\hat{\bm x}\,,\notag
\end{eqnarray}
each with a spectral width $2/\Delta_c$. 

In contrast to the numerical study by Zhang {\it et al.}
\cite{Opt.Lett._36_4569_2011_Zhang}, where the $\bm p\cdot\bm A$
interaction is used and there is no
coupling between different $\bm k$ points, our SBE, which is based on
the $\bm r\cdot\bm E$ interaction, involve derivatives
of the single-particle density matrix with respect to $\bm k$. In the
numerical calculations, we divide the Brillouin zone (BZ) into an
$M\times M$ homogeneous grid, and discretize the derivative in
Eq.~(\ref{eq:removegauge}) as \cite{Phys.Rev.B_56_12847_1997_Marzari} 
\begin{equation}
  \left.\bm \nabla_{\bm q} F(\bm q)\right|_{\bm q=0} \approx
  \frac{a_0^2M^2}{16\pi^2} \sum_{i}\frac{\bm q_i}{M} F\left(\frac{\bm q_i}{M}\right)\,,
\end{equation}
where $\bm q_i$ are chosen as six symmetric points of the honeycomb
lattice 
\begin{equation}
  \{ \bm b_1, \bm b_2, -\bm b_1,
  -\bm b_2, \bm b_1+\bm b_2,
  -(\bm b_1+\bm b_2)\}\,,\notag
\end{equation}

Throughout this work, we are interested in the optical 
response at different frequencies and its dependence on the electric field amplitude
$E_0$, the chemical potential $\mu$ and the gap parameter 
$\Delta$. Other parameters used in the simulation are fixed as $T=300$~K,
$\Delta_c=100$~fs, $\hbar\omega_c=0.6$~eV, and
$\Gamma_i=\Gamma_e=33$~meV. The discrete $\bm k$ points are taken 
from a grid with $M=1500$, and included in the calculation if $\varepsilon_{+\bm
  k}<3.5$~eV; tests involving the inclusion of more $\bm k$ points
confirm that such a restriction leads to converged
numerical simulations.  The time evolution of
Eqs.~(\ref{eq:kbelinear})
and (\ref{eq:kbenonlinear}) is solved by a fourth order Runge-Kutta method
with a time step $\Delta t=0.05$~fs. The current in Eq.~(\ref{eq:j})
is numerically calculated by summing all band indices and all the
effective $\bm k$ points on the grid with an equal weight. After discretization, Eqs.~(\ref{eq:kbelinear})
and (\ref{eq:kbenonlinear}) become linear differential equations for
which the accuracy of the numerical solution is only limited by the
time step. We point out that the density
matrix $\rho_{\bm k}(t)$ acquires a phase dependence on $\bm k$ that changes
with time. At long
enough times $\rho_{\bm k}(t)$ can be strongly dependent on $\bm k$, and then
an accurate calculation of the current from Eq. (\ref{eq:j}) requires a very dense
grid, without which the nonlinear current is buried in numerical noise.
Similarly, a dense grid is also required if the relaxation parameters $\Gamma_{i/e}$
are very small. However, when making calculations for the pulses and
relaxation parameters we adopt here, we find that the nonlinear current can be
determined reliably by the use of the moderate grid identified above.
\begin{figure}[h]
\centering
\includegraphics[width=7cm]{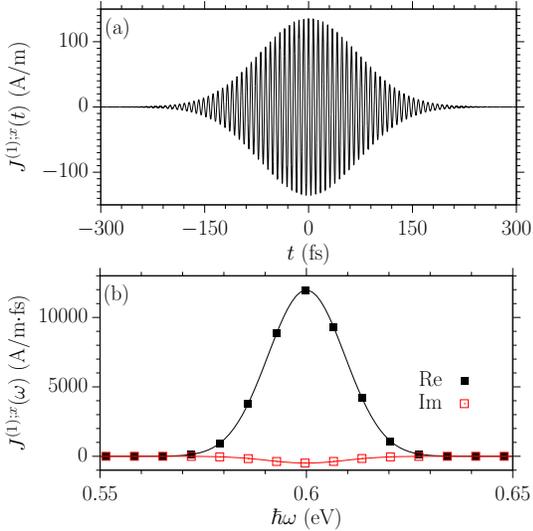}
\caption{(Color online) Linear optical current (a) $J^{(1);x}(t)$
  (b) $J^{(1);x}(\omega)$. The parameters used in the
  calculation are $E_0=10^6$~V/m, $\Delta=0.10$~eV, and
  $\mu=0$. In (b), squares are numerical results, while the curves are fitted to
$J^{(1);x}(\omega)=\sigma_l(\omega_c)E^x(\omega)$.}
\label{fig:evolutionlinear}
\end{figure}

We begin by illustrating the fitting procedure used in this work to extract the
coefficients characterizing the optical response, and consider a weak
incident optical pulse with $E_{0}=10^{6}$~V/m, $\Delta =0.10$~eV, and
$\mu=0$. The linear response can be determined by solving Eq.~(\ref{eq:kbelinear}) numerically,
and using the result to construct $\bm{J}^{(1)}(t)$. \ The result is
shown in Fig.~\ref{fig:evolutionlinear}(a) for an incident field in
the $\hat{\bm x}$ direction,
and the Fourier transform,%
\begin{equation*}
J^{(1);x}(\omega )=\int e^{i\omega t}J^{(1);x}(t)dt
\end{equation*}
is numerically determined and shown in
Fig.~\ref{fig:evolutionlinear}(b). Very generally the linear
  response is of the form 
\begin{equation*}
J^{(1);x}(\omega )=\sigma ^{(1);xx}(\omega )E^{x}(\omega )\,,
\end{equation*}
and $\sigma^{(1);xx}(\omega)$ could be extracted directly for $\omega$
around $\omega_c$. Putting $\sigma _{l}(\omega
_{c})\equiv \sigma ^{(1);xx}(\omega _{c})$, the result is $\sigma
_{l}(\omega _{c})=(1.11-0.05i)\sigma _{0}$, with the universal conductivity 
$\sigma _{0}=e^{2}/(4\hbar)$.

\begin{figure}[t]
\centering
\includegraphics[width=6cm]{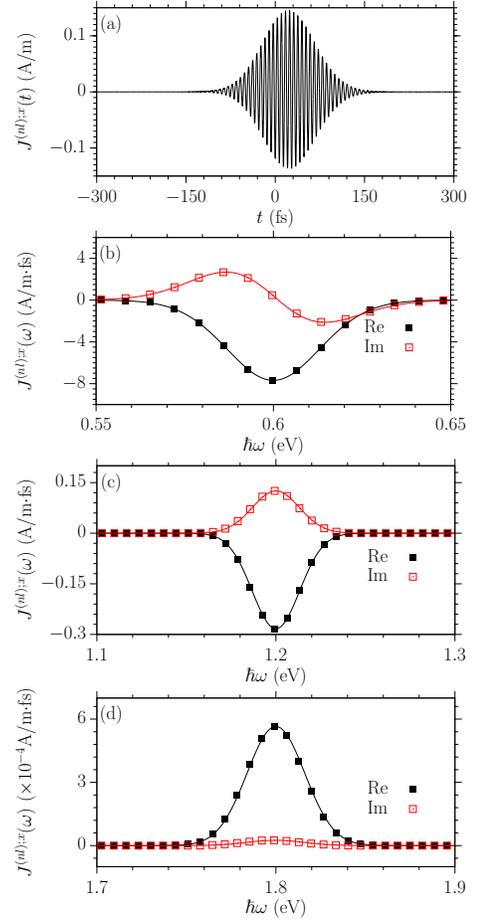}
\caption{(Color online) Nonlinear optical current (a)
  $J^{(nl);x}(t)$, and $J^{(nl);x}(\omega)$ for $\omega$ around (b)
  $\omega_c$, (c) $2\omega_c$, (d) $3\omega_c$. The parameters used in the
  calculation are $E_0=10^6$~V/m, $\Delta=0.10$~eV, and
  $\mu=0$. In figures (b)-(d), squares are numerical results, while
  the curves are fitted to Eq.~(\ref{eq:fitexp}), and the fitting parameters are given in the
  text. } 
\label{fig:evolutionnonlinear}
\end{figure}
The situation is
different for the nonlinear response. It can be determined by solving
Eq.~(\ref{eq:kbenonlinear}) numerically, and using the results to
construct $\bm{J}^{(nl)}(t)$. The result is shown in
Fig.~\ref{fig:evolutionnonlinear}(a); note that it is much smaller than the
linear response, and the peak amplitude is shifted to a time slightly later
than the peak of the linear response. The Fourier transform, 
\begin{equation*}
J^{(nl);x}(\omega )=\int e^{i\omega t}J^{(nl);x}(t)dt,
\end{equation*}
can then be numerically determined. Here we find a significant response
for $\omega$ close to $\omega_{c}$ [corresponding generally to the Kerr
effect and two-photon absorption, and shown in
Fig.~\ref{fig:evolutionnonlinear}(b)], for $\omega$ close
to $2\omega_{c}$ [corresponding to SHG, and shown in
Fig.~\ref{fig:evolutionnonlinear}(c)], and for $\omega $ close to
$3\omega_{c}$ [corresponding to THG, and shown in
Fig.~\ref{fig:evolutionnonlinear}(d)], and of course for $\omega$ 
close to the associated negative frequencies. While
Figs.~\ref{fig:evolutionnonlinear}(c) and \ref{fig:evolutionnonlinear}(d) are essentially Gaussian in form, as was
Fig.~\ref{fig:evolutionlinear}(b), Fig.~\ref{fig:evolutionnonlinear}(b) certainly is not. So
the question arises as to how to characterize the nonlinear response and
identify the relevant response coefficients. 

In the perturbative regime, we can very generally expect a nonlinear response
of the form 
\begin{eqnarray}
  J^{(nl);x}(\omega) &=& \int \frac{d\omega_1}{2\pi} \sigma^{(2);xxx}(\omega_1,\omega-\omega_1)E^x(\omega_1)E^x(\omega-\omega_1)
  \notag\\
  &+& \int \frac{d\omega_1d\omega_2}{(2\pi)^2}
  \sigma^{(3);xxxx}(\omega_1,\omega_2,\omega-\omega_1-\omega_2)\notag\\
&&\times E^x(\omega_1)E^x(\omega_2)E^x(\omega-\omega_1-\omega_2)\,.\label{eq:jpertnon}
\end{eqnarray}
For $\omega _{i,j,k}=\pm \omega _{c}$ and small $\delta _{i,j,k}$, an
approximate analytic perturbation calculation \cite{Phys.Rev.B_91_235320_2015_Cheng} leads to 
\begin{eqnarray}
  \sigma^{(2);xxx}(\omega_i+\delta_1&&, \omega_j+\delta_2) \approx
  s_{1}^{(2)} + \frac{1\text{eV} }{\hbar(\delta_1+\delta_2) + i\gamma } s_{2}^{(2)}\,,\notag\\
\sigma^{(3);xxxx}(\omega_i+\delta_1&&, \omega_j+\delta_2, \omega_k+\delta_3) \approx
  s_{1}^{(3)} \notag\\
&& + \frac{1\text{eV} }{\hbar(\delta_1+\delta_2+\delta_3) + i\gamma } s_{2}^{(3)}\,,\label{eq:fitsigma}
\end{eqnarray}
where $s_{l}^{(2)}$, $s_{l}^{(3)}$, and $\gamma $ are determined in the
calculation, and take on different values for
different choices of the $\omega _{i,j,k}$. Motivated by this, we fit
the results of Fig.~\ref{fig:evolutionnonlinear} by assuming the
conductivity as the form
of Eq.~(\ref{eq:fitsigma}) with taking $s_l^{(2)}$, 
$s_l^{(3)}$, and $\gamma$ as free fitting parameters. This leads to a
fit of the  nonlinear current spectrum around $\Omega =\omega_{c},$ $2\omega_{c}$, and 
$3\omega_{c}$ of the form 
\begin{eqnarray}
  J^{(n);x}(\Omega+\delta) &=& C_{\Omega} \left[s_{1}^{(n)} + \frac{1\text{eV}
    }{\hbar\delta + i\gamma } s_{2}^{(n)}\right]  \notag\\
&&\times
  e^{-(\delta\Delta_c)^2/(4n)}  \frac{\sqrt{\pi}\Delta_c}{\sqrt{n}}E_0^n\,.\label{eq:fitexp}
\end{eqnarray}
Here $n=2$ is used for $\Omega =2\omega _{c}$, and $n=3$ for $\Omega =\omega
_{c}$ or $3\omega _{c}$, with $C_{\Omega }$ describing the permutation
factor relevant for the nonlinear process; $C_{\omega _{c}}=3$, and $%
C_{2\omega _{c}}=C_{3\omega _{c}}=1$. The result of this fitting is shown
by the solid curves in
Fig.~\ref{fig:evolutionnonlinear}(b)-\ref{fig:evolutionnonlinear}(d),
and we can see that indeed a very good 
fit is provided. Once the fit of Eq.~(\ref{eq:fitexp}) is accepted, we
can return to Eq.~(\ref{eq:fitsigma}) and identify the nonlinear
response coefficients $\sigma^{(2);xxx}(\omega _{c},\omega _{c})$
(associated with SHG for a fundamental at $\omega _{c}$), $\sigma ^{(3);xxxx}(\omega
_{c},\omega _{c},\omega _{c})$ (associated with THG for a fundamental
at $\omega _{c}$), and $\sigma ^{(3);xxxx}(-\omega 
_{c},\omega _{c},\omega _{c})$ (associated with the Kerr effect and
two-photon absorption for a fundamental at $\omega _{c}$). For the results
shown in Fig.~\ref{fig:evolutionnonlinear}, for example, we find  $\sigma _{0}^{-1}\sigma ^{(2);xxx}(\omega _{c},\omega
_{c})=(-37.2+16.5i)$~pm/V, with $s_{2}^{(2)}\sim 0$; $\sigma _{0}^{-1}\sigma
^{(3);xxxx}(\omega _{c},\omega _{c},\omega _{c})=(0.91+0.04i)\times
10^{-19}$~m$^{2}$/V$^{2}$ with $s_{2}^{(3)}\sim 0$; and $\sigma _{0}^{-1}\sigma ^{(3);xxxx}(-\omega
_{c},\omega _{c},\omega _{c})=(-4.1+0.2i)\times $ $10^{-16}$~m$^{2}$/V$^{2}$ with 
$\sigma _{0}^{-1}s_{1}^{(3)}=(1463.5+239.7i)\times 10^{-19}$m$^{2}$V$^{2}$, $
\sigma _{0}^{-1}s_{2}^{(3)}=(0.9-202.5i)\times
10^{-19}$~m$^{2}$/V$^{2}$, and $\gamma =36.3$~meV.

For weak incident fields, we can use this strategy to extract coefficients $\sigma ^{(2);xxx}(\omega
_{c},\omega _{c})$, $\sigma ^{(3);xxxx}(\omega _{c},\omega _{c},\omega _{c})$
and $\sigma ^{(3);xxxx}(-\omega _{c},\omega _{c},\omega _{c})$ from our
numerical calculations, confirm that they are independent of the
amplitude $E_{0}$ of the incident field -- as they should be in the
perturbative regime -- and compare them with the results of the approximate but analytic expressions
for these response coefficients. For strong incident fields a strict
perturbative response of the form in Eq.~(\ref{eq:fitsigma}) is not expected to
hold. Still, the nonlinear response can be expected to be characterized by
SHG, THG, and terms that behave
phenomenologically as Kerr and two-photon absorption effects. Thus from
our numerical calculations we can extract an effective $\sigma
^{(2);xxx}(\omega _{c},\omega _{c})$ [which we denote as $\sigma
_{\text{SHG}}(\omega _{c})$], an effective $\sigma ^{(3);xxxx}(\omega _{c},\omega
_{c},\omega _{c})$ [which we denote as $\sigma _{\text{THG}}(\omega _{c})$] and an
effective $\sigma ^{(3);xxxx}(-\omega _{c},\omega _{c},\omega _{c})$ [which
we denote as $\sigma _{nl}(\omega _{c})$]. Unlike the coefficients that
govern the perturbative regime, we can expect the effective coefficients $%
\sigma _{\text{SHG}}(\omega _{c})$, $\sigma _{\text{THG}}(\omega _{c})$, and $\sigma
_{nl}(\omega _{c})$ to depend on the amplitude of the electric field
strength, containing renormalizations of the perturbative response
coefficients in the presence of strong fields. 

Using the fitting scheme described above, we study two examples of the
dependence of the effective conductivities on the chemical potential $\mu$,
the gap parameter $\Delta $, and the electric field amplitude $E_{0}$. In
the first we consider the dependence on $\mu$ and $E_{0}$ with $\Delta =0$,
which we refer to as doped graphene (DG). In the second we consider the
dependence on $\Delta $ and $E_{0}$ with $\mu =0$, which we refer to as
undoped gapped graphene (GG).

\section{Results\label{sec:result}}
\subsection{Comparing numerical calculations to analytic perturbation
  results}
\begin{figure}[t]
\centering
\includegraphics[height=6cm]{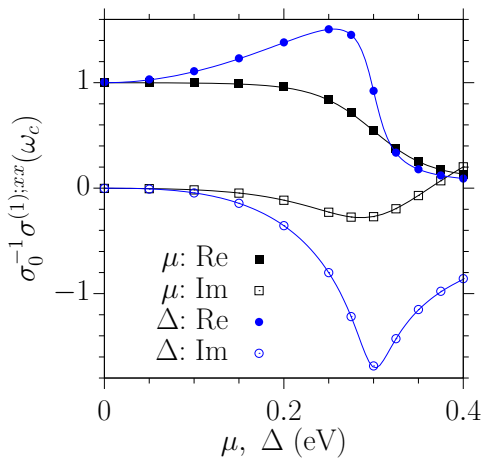}
\caption{(Color online) The linear effective conductivities for DG
  (squares) and GG (dots) for $E_0=10^6$~V/m. The curves are
  calculated from Eq.~(\ref{eq:linearmu}) for DG and Eq.~(\ref{eq:lineardelta}) for GG.}
\label{fig:comparelinear}
\end{figure}
As a benchmark, we first compare the numerical
effective conductivities at a weak electric field $E_{0}=10^{6}$~V/m with
those available from analytic perturbation calculations. We begin with the
linear response. For DG, in previous work \cite{Phys.Rev.B_91_235320_2015_Cheng} we presented the
analytic expression for $\sigma ^{(1);xx}(\omega )$ obtained perturbatively
from the same SBE as Eq. (\ref{eq:krho}), taking into account both interband and
intraband relaxation coefficients $\Gamma _{e}$ and $\Gamma _{i}$
respectively, but using matrix elements and energies correct only 
around the Dirac points; our analytic result is 
\begin{equation}
  \sigma_{\text{DG}}^{(1)xx}(\omega;|\mu|) = \beta
  \int_{-\infty}^{\infty}\!\!\! F_\mu(x, T) [1-F_\mu(x, T)]
  \sigma_{\text{DG};0}^{(1)xx}(\omega; x) dx\,,\label{eq:linearmu}
\end{equation}
where $\beta=1/(k_BT)$ with $k_B$ Boltzmann's constant, and
$F_\mu(x,T)=[1+e^{\beta(x-\mu)}]^{-1}$. The conductivity at zero
temperature is 
\begin{equation*}
\sigma_{\text{DG};0}^{(1);xx}(\omega;\mu)= \frac{i\sigma_0}{\pi}\left\{-{\cal
      G}_{|\mu|}(\hbar\omega+i\Gamma_e)+\frac{4|\mu|}{\hbar\omega+i\Gamma_i}\right\}\,.\notag
\end{equation*}
Here the function ${\cal G}_{|\mu|}(\theta)$ is given for
$\theta=\theta_r + i\theta_i$ as
\begin{eqnarray}
  {\cal G}_{|\mu|}(\theta) && =
  \ln\left|\frac{2|\mu|+\theta}{2|\mu|-\theta}\right| + i \left(\pi
    + \arctan\frac{\theta_r-2|\mu|}{\theta_i} \notag\right.\\
&&\left.-  \arctan\frac{\theta_r+2|\mu|}{\theta_i}\right)\,. \notag
\end{eqnarray}
For GG, because the chemical potential is taken as $0$ and the gap is
nonzero, the net contribution to the linear conductivity from the intraband transitions
(Drude term)  vanishes at zero temperature; even at room temperature
that contribution is negligible, so for GG we can restrict the expression
for the linear conductivity to its interband component,%
\begin{eqnarray}
  \sigma_{\text{GG};\text{inter}}^{(1);xx}(\omega) &=& e^2\sum_{s\bm k}\frac{
    v_{\bar{s}s\bm k}^x \xi^{x}_{s\bar{s}\bm k}(n_{s\bm k}-n_{\bar{s}\bm
      k})}{\hbar\omega-(\varepsilon_{{s}\bm k}-\varepsilon_{\bar{s}\bm
      k})+ i\Gamma_e} \notag\\
  &=&\frac{i\sigma_0}{\pi}\left\{-{\cal
      G}_{\Delta}(\hbar\omega+i\Gamma_e)+\frac{4\Delta}{\hbar\omega+i\Gamma_e}\right\}\notag\\
&&- \frac{i\sigma_0}{\pi}\frac{(2\Delta)^2}{(\hbar\omega+i\Gamma_e)^2}{\cal G}_\Delta(\hbar\omega+i\Gamma_e)\,.\label{eq:lineardelta}
\end{eqnarray}
In Fig.~\ref{fig:comparelinear} we plot the results extracted from our numerical simulations of
Eq.~(\ref{eq:kbelinear}), together with the analytic results in
Eq.~(\ref{eq:linearmu}) and (\ref{eq:lineardelta}) as a function of $\mu$
(for DG) and $\Delta$ (for GG). The agreement is very
good.

\begin{figure*}[t]
\centering
  \includegraphics[height=9.5cm]{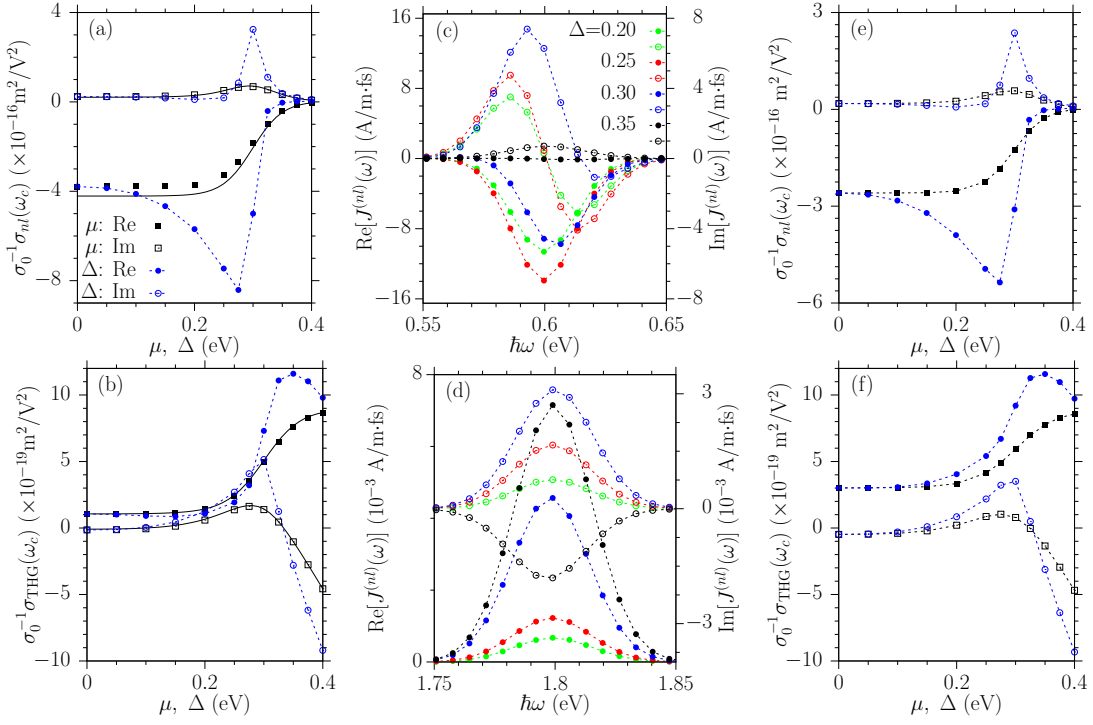}
\caption{(Color online) Nonlinear response for DG and GG: the
  nonlinear conductivity 
  $\sigma_0^{-1}\sigma_{nl}(\omega_c)$ and
  $\sigma_0^{-1}\sigma_{\text{THG}}(\omega_c)$ at (a), (b)
  $E_0=10^6$~V/m  and (e), (f) $2\times 10^7$~V/m;   $J^{(nl)}(\omega)$ of GG with
  $\Delta=0.20$, $0.25$, $0.30$, and $0.35$~eV for $\omega$ around (c)
  $\omega_c$ and  (d) $3\omega_c$. The $y$-axis for the real (imaginary) parts of
  $J^{(nl)}(\omega)$ is at the left (right) hand side of (c) and
  (d).    Solid   curves are calculated from analytic perturbation results
  \cite{Phys.Rev.B_91_235320_2015_Cheng} for DG; dashed
  curves are drawn to guide the eye. }
\label{fig:comparenonlinear}
\end{figure*}
Turning next to the third order response, for the
analytic expressions of
$\sigma^{(3);xxxx}(-\omega_c,\omega_c,\omega_c)$ and
$\sigma^{(3);xxxx}(\omega_c,\omega_c,\omega_c)$ relevant for DG we use
our previous results \cite{Phys.Rev.B_91_235320_2015_Cheng}, including
both interband and intraband relaxation, and with matrix elements and energies taken to be
those that characterize the regions about the Dirac points. For GG with a
nonzero gap parameter, perturbative results for THG were obtained by
Jafari \cite{J.Phys.Condens.Matter._24_205802_2012_Jafari}, but
instead of using the SBE in Eq.~(\ref{eq:krho}) a Kubo formula based
on the ${\bm p}\cdot{\bm A}$ interaction was used, without the inclusion
of any relaxation. Thus while we present our numerical results for $\sigma
_{nl}(\omega _{c})$ and $\sigma _{\text{THG}}(\omega _{c})$ for both DG and GG, we
only compare with the relevant analytic results from perturbation theory
obtained for DG. This is shown in Figs.~\ref{fig:comparenonlinear}(a)
and \ref{fig:comparenonlinear}(b) for $\sigma _{nl}(\omega
_{c})$ and $\sigma _{\text{THG}}(\omega _{c})$ respectively at $E_0=10^6$~V/m. The numerical and analytic results for DG match
very well for chemical potentials over the range shown. There is a noticeable difference between
the numerical and analytically results for $\text{Re}[\sigma_{nl}(\omega
_{c})]$, although it is less than $10\%$, for $\mu<0.3$~eV. We can attribute
this to the singular behavior that $\text{Re}[\sigma _{nl}(\omega _{c})]$
exhibits in the perturbative calculation \cite{Phys.Rev.B_91_235320_2015_Cheng} for $|\mu
| <\hbar \omega _{c}/2$, here $|\mu|<0.3$~eV. Associated with this,
the nonlinear current in the numerical calculation shows a very strong
dependence on the pulse duration and shape, and the strategy
identified above for extracting $\sigma_{nl}(\omega _{c})$ from the
pulse calculation is not completely successful.  

The very good agreement at $E_{0}=10^{6}$~V/m between the effective
conductivities of DG extracted from the numerical calculations, 
and the conductivities predicted by the analytic perturbation theory,
suggests that Eqs.~(\ref{eq:fitsigma}) and (\ref{eq:fitexp}) provide a reasonable fitting procedure, and as
well that  for weak fields the perturbative results presented
earlier \cite{Phys.Rev.B_91_235320_2015_Cheng} are reliable. It also indicates that the usual Dirac point
approximations adopted in the perturbative calculation, involving the linear
dispersion relation and the form of the matrix elements, do not introduce
any significant errors in calculating the linear and nonlinear optical
response of DG at incident photon energies around $\hbar \omega
=0.6$~eV.

We also see from
Figs.~\ref{fig:comparenonlinear}(a) and \ref{fig:comparenonlinear}(b)
that there is a similarity in the dependence of the DG results on
$\mu$ with the dependence of the GG results on $\Delta$. Before turning to the
response of both system at larger field strengths, we address such
similarities in the following section.  

\subsection{Comparing DG and GG}
In investigations of the optical conductivities of doped
graphene, $2|\mu|$ is often treated as an effective gap
\cite{NewJ.Phys._16_53014_2014_Cheng,Phys.Rev.B_91_235320_2015_Cheng}. Since
GG has a real gap of $2\Delta$, it is interesting to compare the dependence of
the optical conductivities on the effective gap $2|\mu|$ induced by
the chemical potential in DG with the real gap $2\Delta $
arising in GG. In linear response, some insight can be gleaned by
comparing the analytic formulas in Eq.~(\ref{eq:linearmu}) and
(\ref{eq:lineardelta}) for DG and GG. In Eq.~(\ref{eq:lineardelta})
the interband velocity matrix elements $\bm v_{+-{\bm k}}$ depend
on $\beta _{\bm \kappa}$, as shown in Table~\ref{tab:app}, and through
that dependence they depend on $\Delta$. If $\beta _{\bm \kappa}$ were not present, only the first
term in the bracket of Eq.~(\ref{eq:lineardelta}) would survive, corresponding to the
interband contribution to the conductivity of DG
\cite{Phys.Rev.B_91_235320_2015_Cheng} with $|\mu|$ replaced by $\Delta$. The presence of $\beta _{\bm   \kappa }$ leads to
the appearance of the other two contributions.  Interestingly, one has the same form as
the Drude term in DG (with $|\mu|$ replaced by $\Delta $), while the
other is new. 

The consequences of the second new term in GG are apparent in
the results shown in Fig.~\ref{fig:comparelinear}; the main differences between the DG and GG
results is that around $\hbar \omega _{c}\approx 2\Delta $ the latter show a
deeper valley in the imaginary part of the conductivity, and a larger peak
in the real part of the conductivity. The real part of the conductivity is
associated with absorption, and through Fermi's Golden Rule it is determined
by both the joint density of states and the velocity matrix elements. Now
for GG the joint density of states for energies around the Dirac points is
given by 
\begin{equation}
{\cal D}(\epsilon) =
2\sum_{\bm k} \delta(\epsilon-(\varepsilon_{+\bm k}+\varepsilon_{-\bm
  k})) = \frac{\epsilon}{2\pi(\hbar v_F)^2}\theta(\epsilon-2\Delta)\,,\notag
\end{equation}
where the factor $2$ comes from the spin degeneracy. Comparing with DG,
the joint density of states is the same for $\epsilon >2|\mu|$ in DG as it is for $\epsilon >2\Delta $ in GG, and so at such
energies the differences in the linear conductivity should be associated
with the velocity matrix elements; and indeed, they can be linked to the
last term in Eq.~(\ref{eq:lineardelta}). 

Turning to the nonlinear response, we first compare the DG and GG results for $\sigma
_{nl}(\omega _{c})$, shown in Fig.~\ref{fig:comparenonlinear}(a). The result for GG shows fine
structure as $\Delta$ is close to $\hbar \omega _{c}/2$. For $\Delta
<\hbar \omega _{c}/2$ both one- and two-photon absorption are present,
and $\text{Re}[\sigma _{nl}(\omega _{c})]$ is negative and increases
in magnitude with increasing $\Delta$. In a manner similar to what is shown by the
results of perturbative calculations
\cite{NewJ.Phys._16_53014_2014_Cheng,Phys.Rev.B_91_235320_2015_Cheng}
at $|\mu|<\hbar \omega _{c}/2$ for DG, we expect that at $\Delta
<\hbar \omega _{c}/2$ for GG the two-photon absorption is associated with saturation as
described at the level of the third-order nonlinearity; it would diverge
when relaxation effects are not included. For $\Delta >\hbar \omega _{c}/2$,
where only two-photon absorption exists, the negative value of
$\text{Re}[\sigma _{nl}(\omega _{c})]$ is induced by the inclusion of
the relaxation \cite{Phys.Rev.B_91_235320_2015_Cheng}. Maximum
absolute values of the imaginary and real parts of
$\sigma_{nl}(\omega_{c})$ occur for GG around $\Delta =\hbar \omega
_{c}/2$, and the differences between the results for GG and DG can again be
attributed to the velocity matrix elements. 

We turn to the results for 
$\sigma _{\text{THG}}(\omega _{c})$ shown in Fig.~\ref{fig:comparenonlinear}(b). The expected similarity of
the results for GG and DG, respectively as a function of $\Delta $ and $|\mu|$,
fails mainly for $\Delta ,\mu >0.25$~eV. Here $\text{Re}\left[ \sigma
  _{\text{THG}}(\omega _{c})\right] $ for GG increases faster than that
of DG, as functions of $\Delta $ and $|\mu|$ respectively, while the
dependences of $\text{Im}[\sigma _{\text{THG}}(\omega _{c})]$ 
for GG and DG are analogous, but with larger absolute values for GG. Again these
differences can be traced back to the different velocity matrix
elements. 

Here we shortly discuss the relation between the fitted effective conductivity
at $\omega_c$ and the amplitude of the optical current calculated from
a laser pulse. As in Eq.~(\ref{eq:fitsigma}), the conductivity shows a 
strong frequency dependence, and thus the value of the conductivity
$\sigma_{nl}(\omega_c)$ at the center frequency of a light pulse is
generally not  a good indication of the amplitude of the optical
response $J^{(nl);x}(\omega)$ if an exciting pulse of light is
actually used. The numerical results for $J^{(nl);x}(\omega)$ are
shown for GG at $\omega$ close to $\omega_c$ in
Fig.~\ref{fig:comparenonlinear}(c), and for $\omega$ close to
$3\omega_c$ in Fig.~\ref{fig:comparenonlinear}(d), for $\Delta=0.20$,
$0.25$, $0.30$, and $0.35$~eV. At $\Delta=\hbar\omega_c/2=0.30$~eV,
both the real and imaginary parts of the nonlinear optical current [black
curves in Fig.~\ref{fig:comparenonlinear}(c)] show a different shape than those at other $\Delta$, although
they do not really exceed them in amplitude. In contrast, there is no obvious shape distortion in the spectrum
shown in Fig.~\ref{fig:comparenonlinear}(d). As such, the
values of $\sigma_{\text{THG}}(\omega_c)$ are consistent with the
magnitude of the optical current of the THG components.  

The results for $\sigma _{nl}(\omega _{c})$ and $\sigma _{\text{THG}}(\omega _{c})$ extracted from
the numerical results for a larger $E_{0}=2\times 10^{7}$~V/m are shown
for both GG and DG in Fig.~\ref{fig:comparenonlinear}(e) and
\ref{fig:comparenonlinear}(f). Note that the dependence of the
effective coefficients of DG on $\left\vert \mu \right\vert $, and
those of GG on $\Delta $, are similar in nature to the dependence of those effective
coefficients at $E_{0}=10^{6}$~V/m, but they take on different
values. Hence we are now beyond the perturbative regime, and cannot
link the effective coefficients $\sigma _{nl}(\omega _{c})$ and
$\sigma _{\text{THG}}(\omega _{c})$ with the perturbative results for
$\sigma ^{(3);xxxx}(-\omega _{c},\omega _{c},\omega _{c})$ and $\sigma ^{(3);xxxx}(\omega _{c},\omega
_{c},\omega _{c})$ respectively. For $\sigma _{nl}(\omega _{c})$ there are
significant differences between the values at $E_{0}=10^{6}$~V/m and
at $E_{0}=2\times 10^{7}$~V/m at all values of $|\mu|$ (or $\Delta $),
while for $\sigma _{\text{THG}}(\omega _{c})$ the differences are
substantial only for $\Delta, |\mu| <0.3$~eV. We
attribute these differences to saturation effects, which we discuss in
the next section.

\subsection{Saturation effects}
We now turn to the dependence of the effective coefficients $\sigma
_{nl}(\omega _{c})$ and $\sigma _{\text{THG}}(\omega _{c})$ on field
strength. We begin with $\sigma_{nl}(\omega_c)$, and note that there
are two different regimes that we can identify for both DG and GG:

(i) $2|\mu|<\hbar\omega_c$ for DG, or $2\Delta<\hbar\omega_c$ for
GG. Here one photon absorption exists and carriers can be injected from the ``$-$'' band to 
the ``$+$'' band. Stronger electric fields inject more carriers.  If the electrons
have a finite lifetime in the states into which they are injected, their injection prevents
the effectiveness of further absorption. Phenomenologically, the effect of the injected carriers on
the total absorption  $\alpha $ is often characterized by
introducing a saturation field strength $E_{\text{sat}}$, 
\begin{equation}
  \alpha = \frac{\alpha_0}{1 + \left({E}/{E_{\text{sat}}}\right)^2}
\end{equation}
with $\alpha_0$ the linear absorption and $E$ the electric
field amplitude in an assumed continuous wave excitation at frequency $\omega $.
For isolated graphene, the absorption of normally incident light is
proportional to $\text{Re}[\sigma _{\text{eff}}^{xx}(\omega )]$, where
$\sigma _{\text{eff}}^{xx}(\omega )$ is a field dependent effective
conductivity, and we would expect
\begin{equation}
  \text{Re}[\sigma_{\text{eff}}^{xx}(\omega)] =
  \frac{\text{Re}[\sigma^{(1);xx}(\omega)]}{1 +
    \left({E}/{E_{\text{sat}}}\right)^2}\,.\label{eq:fitsat}
\end{equation}
However, at weak fields we have \cite{boyd_nonlinearoptics}
\begin{equation}
  \sigma_{\text{eff}}^{xx}(\omega) = \sigma^{(1);xx}(\omega) + 3
  \sigma^{(3);xxxx}(\omega,\omega,-\omega) E^2\,,\label{eq:satsigma}
\end{equation}
where $\sigma^{(3);xxxx}(\omega,\omega,-\omega)$ is the third order
conductivity resulting from a perturbative calculation. Comparing with the weak field
expansion of Eq.~(\ref{eq:fitsat}) we find
\begin{equation}
  E_{\text{sat}}= \sqrt{-\frac{\text{Re}[\sigma^{(1);xx}(\omega)]}{3\text{Re}[\sigma^{(3);xxxx}(-\omega,\omega,\omega)]}}\,.\label{eq:esatpert}
\end{equation}
 For strong electric fields, we assume that Eqs.~(\ref{eq:fitsat}) and (\ref{eq:satsigma})
 work for a field dependent conductivity $\sigma_{nl}(\omega_c)$;
 further, since we extract $\sigma _{nl}(\omega _{c})$ from a
 numerical calculation with the incident field in Eq.~(\ref{eq:elf}) we can identify 
 \begin{equation*}
   \sigma _{\text{eff}}^{xx}(\omega _{c})=\sigma ^{(1);xx}(\omega_{c})+3\sigma
   _{nl}(\omega _{c})E_{0}^{2}\,,
 \end{equation*}
for $\Delta _{c}>\hbar /\Gamma _{i,e}$ where the pulsed excitation
approaches continuous wave excitation, and we can replace $E$ by $E_0$ in
Eq.~(\ref{eq:fitsat}); then we find
\begin{equation}
  \text{Re}[\sigma_{nl}(\omega)]  = -\frac{
    \text{Re}[\sigma^{(1);xx}(\omega)]
  }{3E_{\text{sat}}^2}\frac{1}{1+\left({E_0}/{E_{\text{sat}}}\right)^2}\,. \label{eq:fitsat1}
\end{equation}

\begin{figure}[h]
\centering
\includegraphics[height=6.3cm]{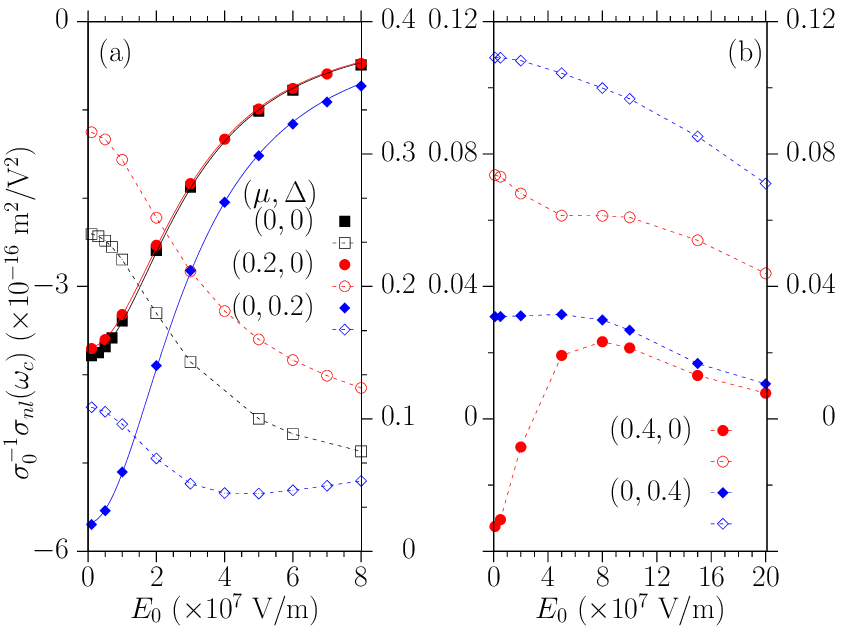} 
\caption{(Color online) Electric field dependence of the nonlinear
  conductivities $\sigma_{nl}(\omega_c)$ for different $(\mu,
  \Delta)$. (a) $(0,0)$ (squares), $(0.2,0)$~eV (circles), and $(0,0.2)$~eV
  (diamonds). (b) $(0.4,0)$ (squares) and $(0,0.4)$~eV (diamonds). The
  real and imaginary parts are given by filled and hollow
  symbols, and their scales are at the left and 
right $y$ axis, respectively. The solid curves in (a) are fitted by functions
$-\frac{\sigma_f}{3E_{\text{sat}}^2}\frac{1}{1+(E_0/E_{\text{sat}})^2}$
with two fitting parameters $\sigma_f$ and $E_{\text{sat}}$, while the dashed
curves are drawn to guide the eye.} 
\label{fig:saturation1}
\end{figure}
In Fig.~\ref{fig:saturation1}(a) we plot the dependence of the numerically determined $\sigma
_{nl}(\omega _{c})$ as a function of $E_{0}$ for three different parameter
sets $(\mu,\Delta)=(0,0)$, $(0.2,0)$, and $(0,0.2)$~eV. The real part
of $\sigma_{nl}(\omega_c)$ is fitted to an expression $-{  \sigma_f}/{[3 (E_{\text{sat}}^2 + E_0^2)]}$
with two parameters $\sigma_f$ and $E_{\text{sat}}$. We find that the fittings (shown as solid
curves) are very good for $\text{Re}[\sigma_f]/\sigma_0\approx-1$, $-1$, and $-1.53$
respectively, with the saturation fields the same in all cases as $E_{\text{sat}}\approx 3\times
10^{7}$~V/m. Comparing
the fitted form with Eq.~(\ref{eq:fitsat1}), and noting that the linear conductivities
are given by $\sigma^{(1);xx}(\omega_c)/\sigma_0=1$,
$0.96-0.11 i$, and $1.38-0.36 i$ respectively
for our parameter sets, the closeness of the fitted $\sigma_f$ with
these linear conductivity $\sigma^{(1);xx}(\omega_c)$ indicates that the saturation can indeed be attributed to linear
absorption. Further, by using the numerical values of $\sigma_{nl}(\omega_c)/\sigma_0$ at a
weak field value $E_0=10^6$~V/m, which for our three parameter sets are $-3.7\times
10^{-16}$, $-3.7\times10^{-16}$, and $-5.7\times
10^{-16}$~m$^2$/V$^2$ respectively, Eq.~(\ref{eq:esatpert}) leads to saturation fields of $3\times 10^7$,
$3\times10^7$, and $2.8\times10^7$~V/m, which are very close to the
fitted values. The field dependence
of $\text{Im}[\sigma _{nl}(\omega _{c})]$, which at least in the weak field
limit can be related to the real part of the nonlinear response via
nonlinear Kramers-Kronig relations, varies in a more complicated way. 

The saturation field can also be estimated from only the linear absorption
coefficients. Physically, the
saturation effect occurs when the injected electron density from one-photon absorption
is comparable to the density of states in the region of $\bm k$ space
where the electrons are injected.  The injected electron density is
$\hbar/\Gamma_i \xi^{xx}(\omega_c)E_m^2$ with the one-photon absorption
coefficients \cite{NewJ.Phys._16_53014_2014_Cheng} 
$\xi^{xx}(\omega_c)=2\text{Re}[\sigma^{(1);xx}(\omega_c)]/(\hbar\omega_c)$
 and the critical field amplitude
$E_m$, while the total available states are estimated as those satisfying $-\Gamma_e\le\varepsilon_{+\bm k}-\varepsilon_{-\bm
  k}-\hbar\omega_c\le\Gamma_e$, which has a density $\int_{\hbar\omega_c-\Gamma_e}^{\hbar\omega_c+\Gamma_e}{\cal
  D}(\epsilon) d\epsilon$. Then
the critical field amplitude $E_m$ is estimated as
\begin{equation}
  E_m \approx \sqrt{\frac{2\Gamma_i\Gamma_e}{\pi } \frac{\sigma_0}{\text{Re}[\sigma^{(1);xx}(\omega_c)]}}\frac{\hbar\omega_c}{\hbar
    |e| v_F }\,.
\end{equation}
This can be used to find approximate values of $E_m\sim 2.8 \times 10^7$~V/m for
those two parameter sets considered for DG, and
$E_m\sim 2.4 \times 10^7$~V/m for the parameter set considered for GG. Both values are close to
the fitted saturation field.  

(ii) $2|\mu|>\hbar\omega_c$ for DG, or $2\Delta>\hbar\omega_c$ for GG. Here
we focus on the frequency regimes $2|\mu|>\hbar\omega_c>|\mu|$
or $2\Delta>\hbar\omega_c>\Delta$ where two photon absorption exists. 
Two photon absorption can inject carriers, but it is less efficient
than one photon absorption. Thus saturation requires 
higher electric fields, and Eq.~(\ref{eq:fitsat}) does not correctly
describe the physics, as shown in
Fig.~\ref{fig:saturation1}(b) for two parameter sets
$(\mu,\Delta)=(0.4,0)$ and $(0,0.4)$~eV, which has different
tendencies compared to the curves in Fig.~\ref{fig:saturation1}(a). For the electric field up
to $E_0=2\times 10^8$~V/m, the imaginary part of $\sigma_{nl}(\omega_c)$ does not
change much for either of these examples. The real part of $\sigma_{nl}(\omega_c)$
of DG changes from negative values to positive values around
$E_0\sim 4\times 10^7$~V/m; while that of GG remains positive and
decreases. For photon energies where even two photon
absorption is absent, we believe that saturation can only occur for
much higher electric fields. 

\begin{figure}[h]
\centering
\includegraphics[height=6cm]{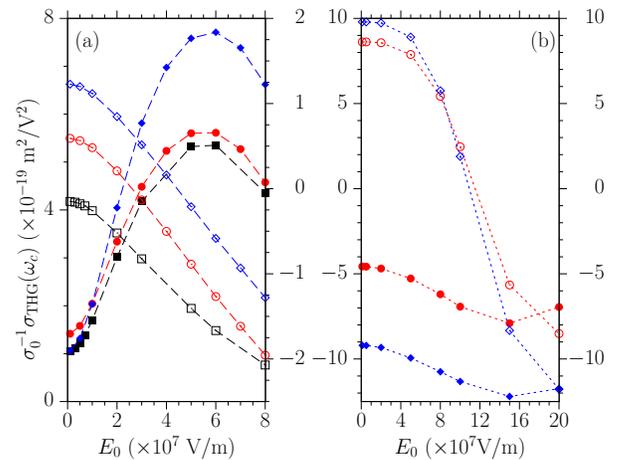}
\caption{(Color online) Electric field dependence of the nonlinear
  conductivities $\sigma_{\text{THG}}(\omega_c)$ for different $(\mu,
  \Delta)$. (a) $(0,0)$ (squares), $(0.2,0)$~eV (circles), and $(0,0.2)$~eV
  (diamonds). (b) $(0.4,0)$ (squares) and $(0,0.4)$~eV (diamonds). The
  real and imaginary parts are given by filled and hollow
  symbols, and their scales are at the left and 
right $y$ axis, respectively. The dashed
curves are drawn to guide the eye.}
\label{fig:saturation2}
\end{figure}

We now turn from $\sigma _{nl}(\omega _{c})$ to $\sigma
_{\text{THG}}(\omega _{c})$. We find that saturation can significantly affect THG, as
shown in Figs.~\ref{fig:saturation2}(a) and
\ref{fig:saturation2}(b). Here again the regimes (i) and (ii)
identified above are relevant. For the results shown in
Fig.~\ref{fig:saturation2}(a) we are in regime (i), where both one- and two-photon
absorption are present. Here both the real and imaginary parts
of $\sigma_{\text{THG}}(\omega_c)$ depend strongly on the electric
field. The imaginary part even changes its sign from positive to negative
values with increasing the electric field, while the real part shows peaks
around $E_0=5\times 10^7$~V/m. Compared to the values at
$E_0=10^6$~V/m, these peak absolute values are about 5 times larger. In 
the measurement of THG\cite{ACSNano_7_8441_2013_Saeynaetjoki,Phys.Rev.B_87_121406_2013_Kumar,Phys.Rev.X_3_021014_2013_Hong}
the procedure used to prepare the samples would indicate the chemical
potentials should be very low; thus saturation may well occur and the
effective THG coefficients $\sigma_{\text{THG}}(\omega_c)$ may be
above their perturbative values. For the results shown in
Fig.~\ref{fig:saturation2}(b) we are in regime (ii), where one-photon absorption is absent
but two-photon absorption is still present. Here the real parts of the $\sigma_{\text{THG}}(\omega_c)$ weakly depend
on the electric field; however, their imaginary parts still strongly depend
on the electric field and change the sign at about
$E_0=10^8$~V/m. 

\begin{figure}[h]
\centering
\includegraphics[height=5cm]{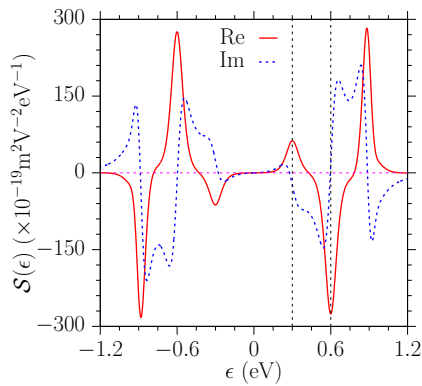}
\caption{(Color online) ${\cal S}(\epsilon)$ at
  $\hbar\omega_c=0.6$~eV, $T=300$~K, and
  $\Gamma_i=\Gamma_e=33$~meV. Those two vertical dashed lines are at
  $\epsilon=0.3$ and $0.6$~eV respectively.}
\label{fig:illustrateb}
\end{figure}
To qualitatively understand how the saturation affects THG, we construct a function 
\begin{equation}
  {\cal S}(\epsilon) = \sigma_0^{-1}\frac{d}{d\epsilon}\sigma^{(3);xxxx}(\omega_c,\omega_c,\omega_c;|\epsilon|)\,,
\end{equation}
Here $\sigma^{(3);xxxx}(\omega_c,\omega_c,\omega_c; |\mu|)$ is the
analytic perturbative third order conductivity of
DG\cite{Phys.Rev.B_91_235320_2015_Cheng} at zero temperature, with the
chemical potential dependence explicitly shown; ${\cal S}(\epsilon)$ describes the contribution of the
electron states at energy $\epsilon$ to the THG. 
For our calculation parameters, $\hbar\omega_c=0.6$~eV, $T=300$~K, and
$\Gamma_i=\Gamma_e=33$~meV, the $\epsilon$ dependence of ${\cal S}(\epsilon)$ is shown in Fig.~\ref{fig:illustrateb}. For a
given $\epsilon$, $\int_{\epsilon-\delta}^{\epsilon+\delta} {\cal
  S}(E) dE $ is the contribution to the THG of the electrons
distributed in the energy range $[\epsilon-\delta,
\epsilon+\delta]$. When the saturation is induced by the one-photon
absorption, the electrons are injected into states with energy
around $\hbar\omega_c/2$ from states with energy around
$-\hbar\omega_c/2$. The contribution of the
population changes to the THG is approximately 
$\propto \left[{\cal
    S}(\hbar\omega_c/2)-{\cal S}(-\hbar\omega_c/2)\right]E_0^2$, with
$E_0^2$ originating from the one-photon injection carrier density. Similarly, the
carriers injected by two-photon absorption contribute $\propto \left[{\cal
  S}(\hbar\omega_c)-{\cal S}(-\hbar\omega_c)\right]E_0^4$, with $E_0^4$
originating from the two-photon injection carrier density. Figure~\ref{fig:illustrateb} shows the real parts of these two terms are positive
and negative respectively. Thus they give competing contributions. For
the results in Fig.~\ref{fig:saturation2}(a), at
small $E_0$, one-photon absorption dominates, and
$\text{Re}[\sigma_{\text{THG}}(\omega_c)]$ increases with $E_0$; at
high $E_0$, two-photon absorption starts to play a role, and the
appearance of a peak of $\text{Re}[\sigma_{\text{THG}}(\omega_c)]$ is
possible. The imaginary part and the results shown in
Fig.~\ref{fig:saturation2}(b) can also be understood in
the same way. 

\subsection{Second harmonic generation}
\begin{figure}[t]
\centering
\includegraphics[height=6cm]{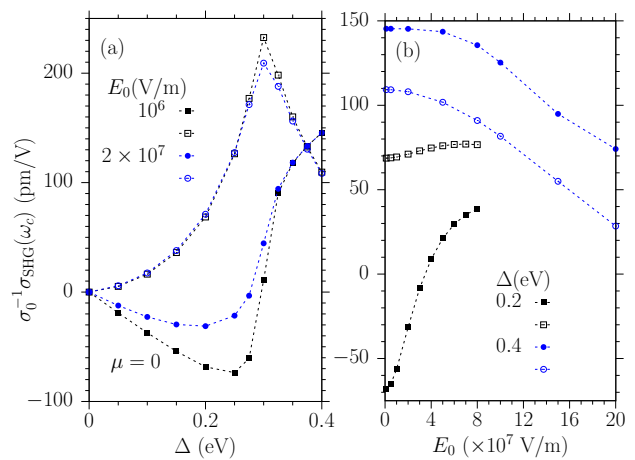}
\caption{(Color online) (a) $\Delta$ dependence of $\sigma_{\text{SHG}}(\omega_c)$ in GG at different
  electric fields $E_0=10^6$~V/m (squares) and $2\times 10^7$~V/m (circles),
  (b) electric field dependence of $\sigma_{\text{SHG}}(\omega_c)$ at different $\Delta=0.2$~eV (squares) and $\Delta=0.4$~eV
  (circles). The real and imaginary parts are given by filled and
  hollow symbols respectively. The dashed lines are drawn to guide the eye.}
\label{fig:shg}
\end{figure}
Finally, we consider the dependence of SHG in GG on the band gap and
electric field amplitude. In parallel with our strategy
for the third order response, we introduce an effective second-order
nonlinear conductivity $\sigma _{\text{SHG}}(\omega _{c})$ which is given by $%
\sigma ^{(2);xxx}(\omega _{c},\omega _{c})$ in the weak field limit, and
extracted for larger fields from the numerical calculations as sketched in
section III. Our results are shown in Fig.~\ref{fig:shg}(a)
and \ref{fig:shg}(b). As expected, a
nonzero $\Delta $, associated with the lack of centre-of-inversion symmetry,
leads to a nonzero SHG response. As $\Delta$ is increased 
from $0$ to $0.4$~eV, the real part of
$\sigma_0^{-1}\sigma_{\text{SHG}}(\omega_c)$ decreases from $0$ to a negative
minimum value (about $-70$~pm/V for $E_0=10^6$~V/m and $-30$~pm/V for
$E_0=2\times10^7$~V/m) around $\Delta=0.25$~eV, then changes sign around
$\Delta=0.3$~eV and reaches a value about $140$~pm/V at
$\Delta=0.4$~eV; they show a strong electric field dependence around
the minimum values.  The imaginary part of $\sigma _{0}^{-1}\sigma
_{\text{SHG}}(\omega _{c})$ has positive
values with a peak $\sim200$~pm/V around $\Delta=0.3$~eV for both electric
field amplitudes considered. Physically,
$\sigma_{\text{SHG}}(\omega_c)$ vanishes as $\Delta=0$, where the
centre-of-inversion symmetry is present, and as $\Delta\to\infty$; for
large $\Delta$ it vanishes as \cite{boyd_nonlinearoptics}  $\propto \Delta^{-4}$. Therefore the existence of a maximum
 of the magnitude of $\sigma_{\text{SHG}}(\omega_c)$ as $\Delta$ is
 increased is not surprising.

To focus on the electric field dependence of $\sigma
_{\text{SHG}}(\omega _{c})$, we plot that dependence in Fig.~\ref{fig:shg}(b) for two gap
parameters, $\Delta =0.2$~eV and $\Delta =0.4$~eV. The real part of
$\sigma_{\text{SHG}}(\omega_c)$ at $\Delta=0.2$~eV shows a strong dependence. It
changes its sign from negative to positive as the electric field
increases from $0$ to $8\times 10^7$~V/m.  The reason is similar to 
the electric field dependence of the third order conductivities, and 
is induced by the saturation effects. However, the imaginary part
changes little over the same range of the electric field.  For
$\Delta=0.4$~eV, where the saturation effects can be ignored, both the
real and imaginary parts  show minor changes up to a electric field $20\times 10^7$~V/m. 

Similar to the estimation for the effective third order
  susceptibilities\cite{NewJ.Phys._16_53014_2014_Cheng,Phys.Rev.B_91_235320_2015_Cheng},
  we calculate the magnitude of the susceptibility of SHG in GG, by
  employing the conversion of $\chi^{(2)}(\omega_c)\approx
\sigma_{\text{SHG}}(\omega_c)/(-2i\omega_c \epsilon_0 d_{\text{gr}})$
with the effective thickness of graphene
$d_{\text{gr}}=3.3$~\mbox{\AA}.  For maximum values of
$|\sigma_0^{-1}\sigma_{\text{SHG}}|\sim 200$~pm/V around
$\Delta=0.3$~eV, we get $\chi^{(2)}\sim 2300$~pm/V. This value is
about 30 times higher than the widely used AgGaSe$_2$
crystal value $68$~pm/V at the same photon
energy\cite{NanoLett._12_2032_2012_Wu}, or a few times larger than
that of monolayer BN, which has a much larger band gap \cite{Phys.Rev.B_72_075416_2005_Guo,Phys.Rev.B_89_081102R_2014_Gruening}.

\section{Conclusion and discussion\label{sec:con}}
In this work, we numerically solved the semiconductor Bloch equations, including
phenomenological relaxation times, for the excitation of both doped and
gapped graphene excited by a pump pulse, and extracted the effective optical 
nonlinear conductivities for second harmonic generation, the Kerr effects,
and third harmonic generation for a given fundamental photon energy
$\hbar\omega_c=0.6$~eV. We focused  
on the dependence of these nonlinear coefficients on the chemical
potential $\mu$ for doped graphene, the gap parameter $\Delta$ for
gapped graphene,
and the electric field amplitude for both. We obtained the following
results. 

(1) For doped graphene: At weak electric fields, all extracted
conductivities (both linear and nonlinear) are in good agreement with the
perturbation results, which is a strong evidence of the correctness of
both the numerical and perturbation calculations. The numerical
results also confirm that both the linear dispersion approximation and
the consideration of only optical transitions around the Dirac points are physically
appropriate in the perturbation calculation with using the standard $\bm r\cdot\bm E$
interaction\footnote{The situation may be different for calculations
  using  $\bm p\cdot\bm A$ interaction. In our numerical calculation
  of linear response with $\bm p\cdot\bm A$ interaction, we find that
  the inclusion of all $\bm k$ in the whole  Brillouin zone is
  necessary for the  imaginary part of the linear conductivity, even
  though the inclusion of $\bm k$ only with transition  energy close
  to the photon energy is adequate for its real part.}. 
With an increase in the electric field amplitude, the effective Kerr coefficient
shows a dependence on the field strength, which can be attributed to 
saturation effects. For $\hbar\omega_c>2|\mu|$ where one-photon
absorption exists, the saturation effects can be characterized by a 
saturation field, which for our relaxation parameters takes a value of
about $3\times 10^7$~V/m. The amplitude of the effective third harmonic generation coefficient can increase up
to 5 times as the electric field changes from $10^6$~V/m to $8\times 10^7$~V/m. However, compared to the two
orders of magnitude difference between the values from the perturbation calculation
and experiments \cite{NewJ.Phys._16_53014_2014_Cheng,Phys.Rev.B_91_235320_2015_Cheng},
this small increment indicates that other effects, such as the
consequences of including more realistic scattering and many-body phenomena, may be important. 

(2) For gapped graphene: The third-order optical conductivity for both
Kerr effect and third harmonic generation in gapped graphene shows obvious peaks or valleys in its $\Delta$
dependence, which is different from the $|\mu|$
dependence in doped graphene due to the nature of the velocity matrix
elements. The susceptibility of second harmonic generation in gapped
graphene is of the order of $10^3$~pm/V, and 
shows a complicated dependence on the gap parameter $\Delta$. 
Compared to the current induced second harmonic generation in doped
graphene, which could be as high as $10^4$~pm/V at similar photon energies under appropriate
conditions\cite{Opt.Express_22_15868_2014_Cheng}, the second harmonic
generation coefficients obtained here are smaller but not that
much. Therefore gapped graphene may also be useful in providing a
second harmonic generation functionality in optical devices.

\acknowledgments
This work has been supported by the EU-FET grant GRAPHENICS (618086),
by the ERC-FP7/2007-2013 grant 336940, by the FWO-Vlaanderen project
G.A002.13N, by the Natural Sciences and Engineering Research Council of
Canada, by VUB-Methusalem, VUB-OZR, and IAP-BELSPO under grant IAP
P7-35.


\begin{thebibliography}{45}%
\makeatletter
\providecommand \@ifxundefined [1]{%
 \@ifx{#1\undefined}
}%
\providecommand \@ifnum [1]{%
 \ifnum #1\expandafter \@firstoftwo
 \else \expandafter \@secondoftwo
 \fi
}%
\providecommand \@ifx [1]{%
 \ifx #1\expandafter \@firstoftwo
 \else \expandafter \@secondoftwo
 \fi
}%
\providecommand \natexlab [1]{#1}%
\providecommand \enquote  [1]{``#1''}%
\providecommand \bibnamefont  [1]{#1}%
\providecommand \bibfnamefont [1]{#1}%
\providecommand \citenamefont [1]{#1}%
\providecommand \href@noop [0]{\@secondoftwo}%
\providecommand \href [0]{\begingroup \@sanitize@url \@href}%
\providecommand \@href[1]{\@@startlink{#1}\@@href}%
\providecommand \@@href[1]{\endgroup#1\@@endlink}%
\providecommand \@sanitize@url [0]{\catcode `\\12\catcode `\$12\catcode
  `\&12\catcode `\#12\catcode `\^12\catcode `\_12\catcode `\%12\relax}%
\providecommand \@@startlink[1]{}%
\providecommand \@@endlink[0]{}%
\providecommand \url  [0]{\begingroup\@sanitize@url \@url }%
\providecommand \@url [1]{\endgroup\@href {#1}{\urlprefix }}%
\providecommand \urlprefix  [0]{URL }%
\providecommand \Eprint [0]{\href }%
\providecommand \doibase [0]{http://dx.doi.org/}%
\providecommand \selectlanguage [0]{\@gobble}%
\providecommand \bibinfo  [0]{\@secondoftwo}%
\providecommand \bibfield  [0]{\@secondoftwo}%
\providecommand \translation [1]{[#1]}%
\providecommand \BibitemOpen [0]{}%
\providecommand \bibitemStop [0]{}%
\providecommand \bibitemNoStop [0]{.\EOS\space}%
\providecommand \EOS [0]{\spacefactor3000\relax}%
\providecommand \BibitemShut  [1]{\csname bibitem#1\endcsname}%
\let\auto@bib@innerbib\@empty
\bibitem [{\citenamefont
  {Mikhailov}(2007)}]{Europhys.Lett._79_27002_2007_Mikhailov}%
  \BibitemOpen
  \bibfield  {author} {\bibinfo {author} {\bibfnamefont {S.~A.}\ \bibnamefont
  {Mikhailov}},\ }\href {http://stacks.iop.org/0295-5075/79/i=2/a=27002}
  {\bibfield  {journal} {\bibinfo  {journal} {Europhys. Lett.}\ }\textbf
  {\bibinfo {volume} {79}},\ \bibinfo {pages} {27002} (\bibinfo {year}
  {2007})}\BibitemShut {NoStop}%
\bibitem [{\citenamefont {Glazov}\ and\ \citenamefont
  {Ganichev}(2014)}]{Phys.Rep._535_101_2014_Glazov}%
  \BibitemOpen
  \bibfield  {author} {\bibinfo {author} {\bibfnamefont {M.}~\bibnamefont
  {Glazov}}\ and\ \bibinfo {author} {\bibfnamefont {S.}~\bibnamefont
  {Ganichev}},\ }\href {\doibase 10.1016/j.physrep.2013.10.003} {\bibfield
  {journal} {\bibinfo  {journal} {Phys. Rep.}\ }\textbf {\bibinfo {volume}
  {535}},\ \bibinfo {pages} {101} (\bibinfo {year} {2014})}\BibitemShut
  {NoStop}%
\bibitem [{\citenamefont {Cheng}\ \emph
  {et~al.}(2014{\natexlab{a}})\citenamefont {Cheng}, \citenamefont
  {Vermeulen},\ and\ \citenamefont {Sipe}}]{NewJ.Phys._16_53014_2014_Cheng}%
  \BibitemOpen
  \bibfield  {author} {\bibinfo {author} {\bibfnamefont {J.~L.}\ \bibnamefont
  {Cheng}}, \bibinfo {author} {\bibfnamefont {N.}~\bibnamefont {Vermeulen}}, \
  and\ \bibinfo {author} {\bibfnamefont {J.~E.}\ \bibnamefont {Sipe}},\ }\href
  {\doibase 10.1088/1367-2630/16/5/053014} {\bibfield  {journal} {\bibinfo
  {journal} {New J. Phys.}\ }\textbf {\bibinfo {volume} {16}},\ \bibinfo
  {pages} {053014} (\bibinfo {year} {2014}{\natexlab{a}})}\BibitemShut
  {NoStop}%
\bibitem [{\citenamefont {Hendry}\ \emph {et~al.}(2010)\citenamefont {Hendry},
  \citenamefont {Hale}, \citenamefont {Moger}, \citenamefont {Savchenko},\ and\
  \citenamefont {Mikhailov}}]{Phys.Rev.Lett._105_097401_2010_Hendry}%
  \BibitemOpen
  \bibfield  {author} {\bibinfo {author} {\bibfnamefont {E.}~\bibnamefont
  {Hendry}}, \bibinfo {author} {\bibfnamefont {P.~J.}\ \bibnamefont {Hale}},
  \bibinfo {author} {\bibfnamefont {J.}~\bibnamefont {Moger}}, \bibinfo
  {author} {\bibfnamefont {A.~K.}\ \bibnamefont {Savchenko}}, \ and\ \bibinfo
  {author} {\bibfnamefont {S.~A.}\ \bibnamefont {Mikhailov}},\ }\href {\doibase
  10.1103/PhysRevLett.105.097401} {\bibfield  {journal} {\bibinfo  {journal}
  {Phys. Rev. Lett.}\ }\textbf {\bibinfo {volume} {105}},\ \bibinfo {pages}
  {097401} (\bibinfo {year} {2010})}\BibitemShut {NoStop}%
\bibitem [{\citenamefont {Bonaccorso}\ \emph {et~al.}(2010)\citenamefont
  {Bonaccorso}, \citenamefont {Sun}, \citenamefont {Hasan},\ and\ \citenamefont
  {Ferrari}}]{Nat.Photon._4_611_2010_Bonaccorso}%
  \BibitemOpen
  \bibfield  {author} {\bibinfo {author} {\bibfnamefont {F.}~\bibnamefont
  {Bonaccorso}}, \bibinfo {author} {\bibfnamefont {Z.}~\bibnamefont {Sun}},
  \bibinfo {author} {\bibfnamefont {T.}~\bibnamefont {Hasan}}, \ and\ \bibinfo
  {author} {\bibfnamefont {A.~C.}\ \bibnamefont {Ferrari}},\ }\href
  {http://dx.doi.org/10.1038/nphoton.2010.186} {\bibfield  {journal} {\bibinfo
  {journal} {Nat. Photon.}\ }\textbf {\bibinfo {volume} {4}},\ \bibinfo {pages}
  {611} (\bibinfo {year} {2010})}\BibitemShut {NoStop}%
\bibitem [{\citenamefont {Gu}\ \emph {et~al.}(2012)\citenamefont {Gu},
  \citenamefont {Petrone}, \citenamefont {McMillan}, \citenamefont {van~der
  Zande}, \citenamefont {Yu}, \citenamefont {Lo}, \citenamefont {Kwong},
  \citenamefont {Hone},\ and\ \citenamefont
  {Wong}}]{Nat.Photon._6_554_2012_Gu}%
  \BibitemOpen
  \bibfield  {author} {\bibinfo {author} {\bibfnamefont {T.}~\bibnamefont
  {Gu}}, \bibinfo {author} {\bibfnamefont {N.}~\bibnamefont {Petrone}},
  \bibinfo {author} {\bibfnamefont {J.~F.}\ \bibnamefont {McMillan}}, \bibinfo
  {author} {\bibfnamefont {A.}~\bibnamefont {van~der Zande}}, \bibinfo {author}
  {\bibfnamefont {M.}~\bibnamefont {Yu}}, \bibinfo {author} {\bibfnamefont
  {G.~Q.}\ \bibnamefont {Lo}}, \bibinfo {author} {\bibfnamefont {D.~L.}\
  \bibnamefont {Kwong}}, \bibinfo {author} {\bibfnamefont {J.}~\bibnamefont
  {Hone}}, \ and\ \bibinfo {author} {\bibfnamefont {C.~W.}\ \bibnamefont
  {Wong}},\ }\href {http://dx.doi.org/10.1038/nphoton.2012.147} {\bibfield
  {journal} {\bibinfo  {journal} {Nat. Photon.}\ }\textbf {\bibinfo {volume}
  {6}},\ \bibinfo {pages} {554} (\bibinfo {year} {2012})}\BibitemShut {NoStop}%
\bibitem [{\citenamefont
  {Yamashita}(2011)}]{IEEE_OSAJournalofLightwaveTechnology_30_427_2011_Yamashita}%
  \BibitemOpen
  \bibfield  {author} {\bibinfo {author} {\bibfnamefont {S.}~\bibnamefont
  {Yamashita}},\ }\href {\doibase 10.1109/jlt.2011.2172574} {\bibfield
  {journal} {\bibinfo  {journal} {IEEE/OSA Journal of Lightwave Technology}\
  }\textbf {\bibinfo {volume} {30}},\ \bibinfo {pages} {427} (\bibinfo {year}
  {2011})}\BibitemShut {NoStop}%
\bibitem [{\citenamefont {Bao}\ and\ \citenamefont
  {Loh}(2012)}]{ACSNano_6_3677_2012_Bao}%
  \BibitemOpen
  \bibfield  {author} {\bibinfo {author} {\bibfnamefont {Q.}~\bibnamefont
  {Bao}}\ and\ \bibinfo {author} {\bibfnamefont {K.~P.}\ \bibnamefont {Loh}},\
  }\href {\doibase 10.1021/nn300989g} {\bibfield  {journal} {\bibinfo
  {journal} {ACS Nano}\ }\textbf {\bibinfo {volume} {6}},\ \bibinfo {pages}
  {3677–3694} (\bibinfo {year} {2012})}\BibitemShut {NoStop}%
\bibitem [{\citenamefont {Horiuchi}(2014)}]{Nat.Photon._8_585_2014_Horiuchi}%
  \BibitemOpen
  \bibfield  {author} {\bibinfo {author} {\bibfnamefont {N.}~\bibnamefont
  {Horiuchi}},\ }\href {\doibase 10.1038/nphoton.2014.186} {\bibfield
  {journal} {\bibinfo  {journal} {Nat. Photon.}\ }\textbf {\bibinfo {volume}
  {8}},\ \bibinfo {pages} {585} (\bibinfo {year} {2014})}\BibitemShut {NoStop}%
\bibitem [{\citenamefont {S\"ayn\"atjoki}\ \emph {et~al.}(2013)\citenamefont
  {S\"ayn\"atjoki}, \citenamefont {Karvonen}, \citenamefont {Riikonen},
  \citenamefont {Kim}, \citenamefont {Mehravar}, \citenamefont {Norwood},
  \citenamefont {Peyghambarian}, \citenamefont {Lipsanen},\ and\ \citenamefont
  {Kieu}}]{ACSNano_7_8441_2013_Saeynaetjoki}%
  \BibitemOpen
  \bibfield  {author} {\bibinfo {author} {\bibfnamefont {A.}~\bibnamefont
  {S\"ayn\"atjoki}}, \bibinfo {author} {\bibfnamefont {L.}~\bibnamefont
  {Karvonen}}, \bibinfo {author} {\bibfnamefont {J.}~\bibnamefont {Riikonen}},
  \bibinfo {author} {\bibfnamefont {W.}~\bibnamefont {Kim}}, \bibinfo {author}
  {\bibfnamefont {S.}~\bibnamefont {Mehravar}}, \bibinfo {author}
  {\bibfnamefont {R.~A.}\ \bibnamefont {Norwood}}, \bibinfo {author}
  {\bibfnamefont {N.}~\bibnamefont {Peyghambarian}}, \bibinfo {author}
  {\bibfnamefont {H.}~\bibnamefont {Lipsanen}}, \ and\ \bibinfo {author}
  {\bibfnamefont {K.}~\bibnamefont {Kieu}},\ }\href {\doibase
  10.1021/nn4042909} {\bibfield  {journal} {\bibinfo  {journal} {ACS Nano}\
  }\textbf {\bibinfo {volume} {7}},\ \bibinfo {pages} {8441} (\bibinfo {year}
  {2013})}\BibitemShut {NoStop}%
\bibitem [{\citenamefont {Kumar}\ \emph {et~al.}(2013)\citenamefont {Kumar},
  \citenamefont {Kumar}, \citenamefont {Gerstenkorn}, \citenamefont {Wang},
  \citenamefont {Chiu}, \citenamefont {Smirl},\ and\ \citenamefont
  {Zhao}}]{Phys.Rev.B_87_121406_2013_Kumar}%
  \BibitemOpen
  \bibfield  {author} {\bibinfo {author} {\bibfnamefont {N.}~\bibnamefont
  {Kumar}}, \bibinfo {author} {\bibfnamefont {J.}~\bibnamefont {Kumar}},
  \bibinfo {author} {\bibfnamefont {C.}~\bibnamefont {Gerstenkorn}}, \bibinfo
  {author} {\bibfnamefont {R.}~\bibnamefont {Wang}}, \bibinfo {author}
  {\bibfnamefont {H.-Y.}\ \bibnamefont {Chiu}}, \bibinfo {author}
  {\bibfnamefont {A.~L.}\ \bibnamefont {Smirl}}, \ and\ \bibinfo {author}
  {\bibfnamefont {H.}~\bibnamefont {Zhao}},\ }\href {\doibase
  10.1103/PhysRevB.87.121406} {\bibfield  {journal} {\bibinfo  {journal} {Phys.
  Rev. B}\ }\textbf {\bibinfo {volume} {87}},\ \bibinfo {pages} {121406}
  (\bibinfo {year} {2013})}\BibitemShut {NoStop}%
\bibitem [{\citenamefont {Hong}\ \emph {et~al.}(2013)\citenamefont {Hong},
  \citenamefont {Dadap}, \citenamefont {Petrone}, \citenamefont {Yeh},
  \citenamefont {Hone},\ and\ \citenamefont
  {Osgood}}]{Phys.Rev.X_3_021014_2013_Hong}%
  \BibitemOpen
  \bibfield  {author} {\bibinfo {author} {\bibfnamefont {S.-Y.}\ \bibnamefont
  {Hong}}, \bibinfo {author} {\bibfnamefont {J.~I.}\ \bibnamefont {Dadap}},
  \bibinfo {author} {\bibfnamefont {N.}~\bibnamefont {Petrone}}, \bibinfo
  {author} {\bibfnamefont {P.-C.}\ \bibnamefont {Yeh}}, \bibinfo {author}
  {\bibfnamefont {J.}~\bibnamefont {Hone}}, \ and\ \bibinfo {author}
  {\bibfnamefont {R.~M.}\ \bibnamefont {Osgood}, \bibfnamefont {Jr.}},\ }\href
  {\doibase 10.1103/PhysRevX.3.021014} {\bibfield  {journal} {\bibinfo
  {journal} {Phys. Rev. X}\ }\textbf {\bibinfo {volume} {3}},\ \bibinfo {pages}
  {021014} (\bibinfo {year} {2013})}\BibitemShut {NoStop}%
\bibitem [{\citenamefont {Yang}\ \emph {et~al.}(2011)\citenamefont {Yang},
  \citenamefont {Feng}, \citenamefont {Wang}, \citenamefont {Huang},
  \citenamefont {Chen}, \citenamefont {Wee},\ and\ \citenamefont
  {Ji}}]{NanoLett._11_2622_2011_Yang}%
  \BibitemOpen
  \bibfield  {author} {\bibinfo {author} {\bibfnamefont {H.}~\bibnamefont
  {Yang}}, \bibinfo {author} {\bibfnamefont {X.}~\bibnamefont {Feng}}, \bibinfo
  {author} {\bibfnamefont {Q.}~\bibnamefont {Wang}}, \bibinfo {author}
  {\bibfnamefont {H.}~\bibnamefont {Huang}}, \bibinfo {author} {\bibfnamefont
  {W.}~\bibnamefont {Chen}}, \bibinfo {author} {\bibfnamefont {A.~T.~S.}\
  \bibnamefont {Wee}}, \ and\ \bibinfo {author} {\bibfnamefont
  {W.}~\bibnamefont {Ji}},\ }\href {\doibase 10.1021/nl200587h} {\bibfield
  {journal} {\bibinfo  {journal} {Nano Lett.}\ }\textbf {\bibinfo {volume}
  {11}},\ \bibinfo {pages} {2622} (\bibinfo {year} {2011})}\BibitemShut
  {NoStop}%
\bibitem [{\citenamefont {Zhang}\ \emph {et~al.}(2012)\citenamefont {Zhang},
  \citenamefont {Virally}, \citenamefont {Bao}, \citenamefont {Ping},
  \citenamefont {Massar}, \citenamefont {Godbout},\ and\ \citenamefont
  {Kockaert}}]{Opt.Lett._37_1856_2012_Zhang}%
  \BibitemOpen
  \bibfield  {author} {\bibinfo {author} {\bibfnamefont {H.}~\bibnamefont
  {Zhang}}, \bibinfo {author} {\bibfnamefont {S.}~\bibnamefont {Virally}},
  \bibinfo {author} {\bibfnamefont {Q.}~\bibnamefont {Bao}}, \bibinfo {author}
  {\bibfnamefont {L.~K.}\ \bibnamefont {Ping}}, \bibinfo {author}
  {\bibfnamefont {S.}~\bibnamefont {Massar}}, \bibinfo {author} {\bibfnamefont
  {N.}~\bibnamefont {Godbout}}, \ and\ \bibinfo {author} {\bibfnamefont
  {P.}~\bibnamefont {Kockaert}},\ }\href {\doibase 10.1364/OL.37.001856}
  {\bibfield  {journal} {\bibinfo  {journal} {Opt. Lett.}\ }\textbf {\bibinfo
  {volume} {37}},\ \bibinfo {pages} {1856} (\bibinfo {year}
  {2012})}\BibitemShut {NoStop}%
\bibitem [{\citenamefont {Wu}\ \emph {et~al.}(2011)\citenamefont {Wu},
  \citenamefont {Zhang}, \citenamefont {Yan}, \citenamefont {Bian},
  \citenamefont {Wang}, \citenamefont {Bai}, \citenamefont {Lu}, \citenamefont
  {Zhao},\ and\ \citenamefont {Wang}}]{NanoLett._11_5159_2011_Wu}%
  \BibitemOpen
  \bibfield  {author} {\bibinfo {author} {\bibfnamefont {R.}~\bibnamefont
  {Wu}}, \bibinfo {author} {\bibfnamefont {Y.}~\bibnamefont {Zhang}}, \bibinfo
  {author} {\bibfnamefont {S.}~\bibnamefont {Yan}}, \bibinfo {author}
  {\bibfnamefont {F.}~\bibnamefont {Bian}}, \bibinfo {author} {\bibfnamefont
  {W.}~\bibnamefont {Wang}}, \bibinfo {author} {\bibfnamefont {X.}~\bibnamefont
  {Bai}}, \bibinfo {author} {\bibfnamefont {X.}~\bibnamefont {Lu}}, \bibinfo
  {author} {\bibfnamefont {J.}~\bibnamefont {Zhao}}, \ and\ \bibinfo {author}
  {\bibfnamefont {E.}~\bibnamefont {Wang}},\ }\href {\doibase
  10.1021/nl2023405} {\bibfield  {journal} {\bibinfo  {journal} {Nano Lett.}\
  }\textbf {\bibinfo {volume} {11}},\ \bibinfo {pages} {5159} (\bibinfo {year}
  {2011})}\BibitemShut {NoStop}%
\bibitem [{\citenamefont {Dean}\ and\ \citenamefont {van
  Driel}(2009)}]{Appl.Phys.Lett._95_261910_2009_Dean}%
  \BibitemOpen
  \bibfield  {author} {\bibinfo {author} {\bibfnamefont {J.~J.}\ \bibnamefont
  {Dean}}\ and\ \bibinfo {author} {\bibfnamefont {H.~M.}\ \bibnamefont {van
  Driel}},\ }\href {\doibase 10.1063/1.3275740} {\bibfield  {journal} {\bibinfo
   {journal} {Appl. Phys. Lett.}\ }\textbf {\bibinfo {volume} {95}},\ \bibinfo
  {eid} {261910} (\bibinfo {year} {2009})}\BibitemShut {NoStop}%
\bibitem [{\citenamefont {Dean}\ and\ \citenamefont {van
  Driel}(2010)}]{Phys.Rev.B_82_125411_2010_Dean}%
  \BibitemOpen
  \bibfield  {author} {\bibinfo {author} {\bibfnamefont {J.~J.}\ \bibnamefont
  {Dean}}\ and\ \bibinfo {author} {\bibfnamefont {H.~M.}\ \bibnamefont {van
  Driel}},\ }\href {\doibase 10.1103/physrevb.82.125411} {\bibfield  {journal}
  {\bibinfo  {journal} {Phys. Rev. B}\ }\textbf {\bibinfo {volume} {82}},\
  \bibinfo {pages} {125411} (\bibinfo {year} {2010})}\BibitemShut {NoStop}%
\bibitem [{\citenamefont {Bykov}\ \emph {et~al.}(2012)\citenamefont {Bykov},
  \citenamefont {Murzina}, \citenamefont {Rybin},\ and\ \citenamefont
  {Obraztsova}}]{Phys.Rev.B_85_121413_2012_Bykov}%
  \BibitemOpen
  \bibfield  {author} {\bibinfo {author} {\bibfnamefont {A.~Y.}\ \bibnamefont
  {Bykov}}, \bibinfo {author} {\bibfnamefont {T.~V.}\ \bibnamefont {Murzina}},
  \bibinfo {author} {\bibfnamefont {M.~G.}\ \bibnamefont {Rybin}}, \ and\
  \bibinfo {author} {\bibfnamefont {E.~D.}\ \bibnamefont {Obraztsova}},\ }\href
  {\doibase 10.1103/PhysRevB.85.121413} {\bibfield  {journal} {\bibinfo
  {journal} {Phys. Rev. B}\ }\textbf {\bibinfo {volume} {85}},\ \bibinfo
  {pages} {121413} (\bibinfo {year} {2012})}\BibitemShut {NoStop}%
\bibitem [{\citenamefont {An}\ \emph {et~al.}(2013)\citenamefont {An},
  \citenamefont {Nelson}, \citenamefont {Lee},\ and\ \citenamefont
  {Diebold}}]{NanoLett._13_2104_2013_An}%
  \BibitemOpen
  \bibfield  {author} {\bibinfo {author} {\bibfnamefont {Y.~Q.}\ \bibnamefont
  {An}}, \bibinfo {author} {\bibfnamefont {F.}~\bibnamefont {Nelson}}, \bibinfo
  {author} {\bibfnamefont {J.~U.}\ \bibnamefont {Lee}}, \ and\ \bibinfo
  {author} {\bibfnamefont {A.~C.}\ \bibnamefont {Diebold}},\ }\href {\doibase
  10.1021/nl4004514} {\bibfield  {journal} {\bibinfo  {journal} {Nano Lett.}\
  }\textbf {\bibinfo {volume} {13}},\ \bibinfo {pages} {2104} (\bibinfo {year}
  {2013})}\BibitemShut {NoStop}%
\bibitem [{\citenamefont {An}\ \emph {et~al.}(2014)\citenamefont {An},
  \citenamefont {Rowe}, \citenamefont {Dougherty}, \citenamefont {Lee},\ and\
  \citenamefont {Diebold}}]{Phys.Rev.B_89_115310_2014_An}%
  \BibitemOpen
  \bibfield  {author} {\bibinfo {author} {\bibfnamefont {Y.~Q.}\ \bibnamefont
  {An}}, \bibinfo {author} {\bibfnamefont {J.~E.}\ \bibnamefont {Rowe}},
  \bibinfo {author} {\bibfnamefont {D.~B.}\ \bibnamefont {Dougherty}}, \bibinfo
  {author} {\bibfnamefont {J.~U.}\ \bibnamefont {Lee}}, \ and\ \bibinfo
  {author} {\bibfnamefont {A.~C.}\ \bibnamefont {Diebold}},\ }\href {\doibase
  10.1103/physrevb.89.115310} {\bibfield  {journal} {\bibinfo  {journal} {Phys.
  Rev. B}\ }\textbf {\bibinfo {volume} {89}},\ \bibinfo {pages} {115310}
  (\bibinfo {year} {2014})}\BibitemShut {NoStop}%
\bibitem [{\citenamefont {Lin}\ \emph {et~al.}(2014)\citenamefont {Lin},
  \citenamefont {Weng}, \citenamefont {Lyu}, \citenamefont {Tsai},\ and\
  \citenamefont {Su}}]{Appl.Phys.Lett._105_151605_2014_Lin}%
  \BibitemOpen
  \bibfield  {author} {\bibinfo {author} {\bibfnamefont {K.-H.}\ \bibnamefont
  {Lin}}, \bibinfo {author} {\bibfnamefont {S.-W.}\ \bibnamefont {Weng}},
  \bibinfo {author} {\bibfnamefont {P.-W.}\ \bibnamefont {Lyu}}, \bibinfo
  {author} {\bibfnamefont {T.-R.}\ \bibnamefont {Tsai}}, \ and\ \bibinfo
  {author} {\bibfnamefont {W.-B.}\ \bibnamefont {Su}},\ }\href {\doibase
  10.1063/1.4898065} {\bibfield  {journal} {\bibinfo  {journal} {Appl. Phys.
  Lett.}\ }\textbf {\bibinfo {volume} {105}},\ \bibinfo {pages} {151605}
  (\bibinfo {year} {2014})}\BibitemShut {NoStop}%
\bibitem [{\citenamefont {Sun}\ \emph {et~al.}(2010)\citenamefont {Sun},
  \citenamefont {Divin}, \citenamefont {Rioux}, \citenamefont {Sipe},
  \citenamefont {Berger}, \citenamefont {de~Heer}, \citenamefont {First},\ and\
  \citenamefont {Norris}}]{NanoLett._10_1293_2010_Sun}%
  \BibitemOpen
  \bibfield  {author} {\bibinfo {author} {\bibfnamefont {D.}~\bibnamefont
  {Sun}}, \bibinfo {author} {\bibfnamefont {C.}~\bibnamefont {Divin}}, \bibinfo
  {author} {\bibfnamefont {J.}~\bibnamefont {Rioux}}, \bibinfo {author}
  {\bibfnamefont {J.~E.}\ \bibnamefont {Sipe}}, \bibinfo {author}
  {\bibfnamefont {C.}~\bibnamefont {Berger}}, \bibinfo {author} {\bibfnamefont
  {W.~A.}\ \bibnamefont {de~Heer}}, \bibinfo {author} {\bibfnamefont {P.~N.}\
  \bibnamefont {First}}, \ and\ \bibinfo {author} {\bibfnamefont {T.~B.}\
  \bibnamefont {Norris}},\ }\href {\doibase 10.1021/nl9040737} {\bibfield
  {journal} {\bibinfo  {journal} {Nano Lett.}\ }\textbf {\bibinfo {volume}
  {10}},\ \bibinfo {pages} {1293} (\bibinfo {year} {2010})}\BibitemShut
  {NoStop}%
\bibitem [{\citenamefont {Sun}\ \emph {et~al.}(2012{\natexlab{a}})\citenamefont
  {Sun}, \citenamefont {Divin}, \citenamefont {Mihnev}, \citenamefont {Winzer},
  \citenamefont {Malic}, \citenamefont {Knorr}, \citenamefont {Sipe},
  \citenamefont {Berger}, \citenamefont {de~Heer}, \citenamefont {First},\ and\
  \citenamefont {Norris}}]{NewJ.Phys._14_105012_2012_Sun}%
  \BibitemOpen
  \bibfield  {author} {\bibinfo {author} {\bibfnamefont {D.}~\bibnamefont
  {Sun}}, \bibinfo {author} {\bibfnamefont {C.}~\bibnamefont {Divin}}, \bibinfo
  {author} {\bibfnamefont {M.}~\bibnamefont {Mihnev}}, \bibinfo {author}
  {\bibfnamefont {T.}~\bibnamefont {Winzer}}, \bibinfo {author} {\bibfnamefont
  {E.}~\bibnamefont {Malic}}, \bibinfo {author} {\bibfnamefont
  {A.}~\bibnamefont {Knorr}}, \bibinfo {author} {\bibfnamefont {J.~E.}\
  \bibnamefont {Sipe}}, \bibinfo {author} {\bibfnamefont {C.}~\bibnamefont
  {Berger}}, \bibinfo {author} {\bibfnamefont {W.~A.}\ \bibnamefont {de~Heer}},
  \bibinfo {author} {\bibfnamefont {P.~N.}\ \bibnamefont {First}}, \ and\
  \bibinfo {author} {\bibfnamefont {T.~B.}\ \bibnamefont {Norris}},\ }\href
  {http://stacks.iop.org/1367-2630/14/i=10/a=105012} {\bibfield  {journal}
  {\bibinfo  {journal} {New J. Phys.}\ }\textbf {\bibinfo {volume} {14}},\
  \bibinfo {pages} {105012} (\bibinfo {year} {2012}{\natexlab{a}})}\BibitemShut
  {NoStop}%
\bibitem [{\citenamefont {Sun}\ \emph {et~al.}(2012{\natexlab{b}})\citenamefont
  {Sun}, \citenamefont {Rioux}, \citenamefont {Sipe}, \citenamefont {Zou},
  \citenamefont {Mihnev}, \citenamefont {Berger}, \citenamefont {de~Heer},
  \citenamefont {First},\ and\ \citenamefont
  {Norris}}]{Phys.Rev.B_85_165427_2012_Sun}%
  \BibitemOpen
  \bibfield  {author} {\bibinfo {author} {\bibfnamefont {D.}~\bibnamefont
  {Sun}}, \bibinfo {author} {\bibfnamefont {J.}~\bibnamefont {Rioux}}, \bibinfo
  {author} {\bibfnamefont {J.~E.}\ \bibnamefont {Sipe}}, \bibinfo {author}
  {\bibfnamefont {Y.}~\bibnamefont {Zou}}, \bibinfo {author} {\bibfnamefont
  {M.~T.}\ \bibnamefont {Mihnev}}, \bibinfo {author} {\bibfnamefont
  {C.}~\bibnamefont {Berger}}, \bibinfo {author} {\bibfnamefont {W.~A.}\
  \bibnamefont {de~Heer}}, \bibinfo {author} {\bibfnamefont {P.~N.}\
  \bibnamefont {First}}, \ and\ \bibinfo {author} {\bibfnamefont {T.~B.}\
  \bibnamefont {Norris}},\ }\href {\doibase 10.1103/physrevb.85.165427}
  {\bibfield  {journal} {\bibinfo  {journal} {Phys. Rev. B}\ }\textbf {\bibinfo
  {volume} {85}},\ \bibinfo {pages} {165427} (\bibinfo {year}
  {2012}{\natexlab{b}})}\BibitemShut {NoStop}%
\bibitem [{\citenamefont {Rioux}\ \emph {et~al.}(2011)\citenamefont {Rioux},
  \citenamefont {Burkard},\ and\ \citenamefont
  {Sipe}}]{Phys.Rev.B_83_195406_2011_Rioux}%
  \BibitemOpen
  \bibfield  {author} {\bibinfo {author} {\bibfnamefont {J.}~\bibnamefont
  {Rioux}}, \bibinfo {author} {\bibfnamefont {G.}~\bibnamefont {Burkard}}, \
  and\ \bibinfo {author} {\bibfnamefont {J.~E.}\ \bibnamefont {Sipe}},\ }\href
  {\doibase 10.1103/PhysRevB.83.195406} {\bibfield  {journal} {\bibinfo
  {journal} {Phys. Rev. B}\ }\textbf {\bibinfo {volume} {83}},\ \bibinfo
  {pages} {195406} (\bibinfo {year} {2011})}\BibitemShut {NoStop}%
\bibitem [{\citenamefont {Rioux}\ \emph {et~al.}(2014)\citenamefont {Rioux},
  \citenamefont {Sipe},\ and\ \citenamefont
  {Burkard}}]{Phys.Rev.B_90_115424_2014_Rioux}%
  \BibitemOpen
  \bibfield  {author} {\bibinfo {author} {\bibfnamefont {J.}~\bibnamefont
  {Rioux}}, \bibinfo {author} {\bibfnamefont {J.~E.}\ \bibnamefont {Sipe}}, \
  and\ \bibinfo {author} {\bibfnamefont {G.}~\bibnamefont {Burkard}},\ }\href
  {\doibase 10.1103/physrevb.90.115424} {\bibfield  {journal} {\bibinfo
  {journal} {Phys. Rev. B}\ }\textbf {\bibinfo {volume} {90}},\ \bibinfo
  {pages} {115424} (\bibinfo {year} {2014})}\BibitemShut {NoStop}%
\bibitem [{\citenamefont {Mikhailov}\ and\ \citenamefont
  {Ziegler}(2008)}]{J.Phys.Condens.Matter_20_384204_Mikhailov}%
  \BibitemOpen
  \bibfield  {author} {\bibinfo {author} {\bibfnamefont {S.~A.}\ \bibnamefont
  {Mikhailov}}\ and\ \bibinfo {author} {\bibfnamefont {K.}~\bibnamefont
  {Ziegler}},\ }\href {http://stacks.iop.org/0953-8984/20/i=38/a=384204}
  {\bibfield  {journal} {\bibinfo  {journal} {J. Phys. Condens. Matter}\
  }\textbf {\bibinfo {volume} {20}},\ \bibinfo {pages} {384204} (\bibinfo
  {year} {2008})}\BibitemShut {NoStop}%
\bibitem [{arX()}]{arXiv:1011.4841}%
  \BibitemOpen
  \href@noop {} {}\bibinfo {note} {F. T. Vasko, arXiv:1011.4841
  (2010).}\BibitemShut {Stop}%
\bibitem [{\citenamefont {Zhang}\ and\ \citenamefont
  {Voss}(2011)}]{Opt.Lett._36_4569_2011_Zhang}%
  \BibitemOpen
  \bibfield  {author} {\bibinfo {author} {\bibfnamefont {Z.}~\bibnamefont
  {Zhang}}\ and\ \bibinfo {author} {\bibfnamefont {P.~L.}\ \bibnamefont
  {Voss}},\ }\href {\doibase 10.1364/OL.36.004569} {\bibfield  {journal}
  {\bibinfo  {journal} {Opt. Lett.}\ }\textbf {\bibinfo {volume} {36}},\
  \bibinfo {pages} {4569} (\bibinfo {year} {2011})}\BibitemShut {NoStop}%
\bibitem [{\citenamefont
  {Jafari}(2012)}]{J.Phys.Condens.Matter._24_205802_2012_Jafari}%
  \BibitemOpen
  \bibfield  {author} {\bibinfo {author} {\bibfnamefont {S.~A.}\ \bibnamefont
  {Jafari}},\ }\href {http://stacks.iop.org/0953-8984/24/i=20/a=205802}
  {\bibfield  {journal} {\bibinfo  {journal} {J. Phys. Condens. Matter}\
  }\textbf {\bibinfo {volume} {24}},\ \bibinfo {pages} {205802} (\bibinfo
  {year} {2012})}\BibitemShut {NoStop}%
\bibitem [{\citenamefont {Avetissian}\ \emph
  {et~al.}(2012{\natexlab{a}})\citenamefont {Avetissian}, \citenamefont
  {Avetissian}, \citenamefont {Mkrtchian},\ and\ \citenamefont
  {Sedrakian}}]{Phys.Rev.B_85_115443_2012_Avetissian}%
  \BibitemOpen
  \bibfield  {author} {\bibinfo {author} {\bibfnamefont {H.~K.}\ \bibnamefont
  {Avetissian}}, \bibinfo {author} {\bibfnamefont {A.~K.}\ \bibnamefont
  {Avetissian}}, \bibinfo {author} {\bibfnamefont {G.~F.}\ \bibnamefont
  {Mkrtchian}}, \ and\ \bibinfo {author} {\bibfnamefont {K.~V.}\ \bibnamefont
  {Sedrakian}},\ }\href {\doibase 10.1103/physrevb.85.115443} {\bibfield
  {journal} {\bibinfo  {journal} {Phys. Rev. B}\ }\textbf {\bibinfo {volume}
  {85}},\ \bibinfo {pages} {115443} (\bibinfo {year}
  {2012}{\natexlab{a}})}\BibitemShut {NoStop}%
\bibitem [{\citenamefont {Avetissian}\ \emph
  {et~al.}(2012{\natexlab{b}})\citenamefont {Avetissian}, \citenamefont
  {Avetissian}, \citenamefont {Mkrtchian},\ and\ \citenamefont
  {Sedrakian}}]{J.Nanophoton._6_61702_2012_Avetissian}%
  \BibitemOpen
  \bibfield  {author} {\bibinfo {author} {\bibfnamefont {H.~K.}\ \bibnamefont
  {Avetissian}}, \bibinfo {author} {\bibfnamefont {A.~K.}\ \bibnamefont
  {Avetissian}}, \bibinfo {author} {\bibfnamefont {G.~F.}\ \bibnamefont
  {Mkrtchian}}, \ and\ \bibinfo {author} {\bibfnamefont {K.~V.}\ \bibnamefont
  {Sedrakian}},\ }\href {\doibase 10.1117/1.jnp.6.061702} {\bibfield  {journal}
  {\bibinfo  {journal} {J. Nanophoton.}\ }\textbf {\bibinfo {volume} {6}},\
  \bibinfo {pages} {061702} (\bibinfo {year} {2012}{\natexlab{b}})}\BibitemShut
  {NoStop}%
\bibitem [{\citenamefont {Avetissian}\ \emph
  {et~al.}(2013{\natexlab{a}})\citenamefont {Avetissian}, \citenamefont
  {Mkrtchian}, \citenamefont {Batrakov}, \citenamefont {Maksimenko},\ and\
  \citenamefont {Hoffmann}}]{Phys.Rev.B_88_165411_2013_Avetissian}%
  \BibitemOpen
  \bibfield  {author} {\bibinfo {author} {\bibfnamefont {H.~K.}\ \bibnamefont
  {Avetissian}}, \bibinfo {author} {\bibfnamefont {G.~F.}\ \bibnamefont
  {Mkrtchian}}, \bibinfo {author} {\bibfnamefont {K.~G.}\ \bibnamefont
  {Batrakov}}, \bibinfo {author} {\bibfnamefont {S.~A.}\ \bibnamefont
  {Maksimenko}}, \ and\ \bibinfo {author} {\bibfnamefont {A.}~\bibnamefont
  {Hoffmann}},\ }\href {\doibase 10.1103/physrevb.88.165411} {\bibfield
  {journal} {\bibinfo  {journal} {Phys. Rev. B}\ }\textbf {\bibinfo {volume}
  {88}},\ \bibinfo {pages} {165411} (\bibinfo {year}
  {2013}{\natexlab{a}})}\BibitemShut {NoStop}%
\bibitem [{\citenamefont {Avetissian}\ \emph
  {et~al.}(2013{\natexlab{b}})\citenamefont {Avetissian}, \citenamefont
  {Mkrtchian}, \citenamefont {Batrakov}, \citenamefont {Maksimenko},\ and\
  \citenamefont {Hoffmann}}]{Phys.Rev.B_88_245411_2013_Avetissian}%
  \BibitemOpen
  \bibfield  {author} {\bibinfo {author} {\bibfnamefont {H.~K.}\ \bibnamefont
  {Avetissian}}, \bibinfo {author} {\bibfnamefont {G.~F.}\ \bibnamefont
  {Mkrtchian}}, \bibinfo {author} {\bibfnamefont {K.~G.}\ \bibnamefont
  {Batrakov}}, \bibinfo {author} {\bibfnamefont {S.~A.}\ \bibnamefont
  {Maksimenko}}, \ and\ \bibinfo {author} {\bibfnamefont {A.}~\bibnamefont
  {Hoffmann}},\ }\href {\doibase 10.1103/physrevb.88.245411} {\bibfield
  {journal} {\bibinfo  {journal} {Phys. Rev. B}\ }\textbf {\bibinfo {volume}
  {88}},\ \bibinfo {pages} {245411} (\bibinfo {year}
  {2013}{\natexlab{b}})}\BibitemShut {NoStop}%
\bibitem [{\citenamefont {Cheng}\ \emph
  {et~al.}(2014{\natexlab{b}})\citenamefont {Cheng}, \citenamefont
  {Vermeulen},\ and\ \citenamefont {Sipe}}]{Opt.Express_22_15868_2014_Cheng}%
  \BibitemOpen
  \bibfield  {author} {\bibinfo {author} {\bibfnamefont {J.~L.}\ \bibnamefont
  {Cheng}}, \bibinfo {author} {\bibfnamefont {N.}~\bibnamefont {Vermeulen}}, \
  and\ \bibinfo {author} {\bibfnamefont {J.~E.}\ \bibnamefont {Sipe}},\ }\href
  {\doibase 10.1364/OE.22.015868} {\bibfield  {journal} {\bibinfo  {journal}
  {Opt. Express}\ }\textbf {\bibinfo {volume} {22}},\ \bibinfo {pages} {15868}
  (\bibinfo {year} {2014}{\natexlab{b}})}\BibitemShut {NoStop}%
\bibitem [{\citenamefont
  {Mikhailov}(2014)}]{Phys.Rev.B_90_241301_2014_Mikhailov}%
  \BibitemOpen
  \bibfield  {author} {\bibinfo {author} {\bibfnamefont {S.~A.}\ \bibnamefont
  {Mikhailov}},\ }\href {\doibase 10.1103/physrevb.90.241301} {\bibfield
  {journal} {\bibinfo  {journal} {Phys. Rev. B}\ }\textbf {\bibinfo {volume}
  {90}},\ \bibinfo {pages} {241301(R)} (\bibinfo {year} {2014})}\BibitemShut
  {NoStop}%
\bibitem [{\citenamefont {Cheng}\ \emph {et~al.}(2015)\citenamefont {Cheng},
  \citenamefont {Vermeulen},\ and\ \citenamefont
  {Sipe}}]{Phys.Rev.B_91_235320_2015_Cheng}%
  \BibitemOpen
  \bibfield  {author} {\bibinfo {author} {\bibfnamefont {J.~L.}\ \bibnamefont
  {Cheng}}, \bibinfo {author} {\bibfnamefont {N.}~\bibnamefont {Vermeulen}}, \
  and\ \bibinfo {author} {\bibfnamefont {J.~E.}\ \bibnamefont {Sipe}},\ }\href
  {\doibase http://dx.doi.org/10.1103/PhysRevB.91.235320} {\bibfield  {journal}
  {\bibinfo  {journal} {Phys. Rev. B}\ }\textbf {\bibinfo {volume} {91}},\
  \bibinfo {pages} {235320} (\bibinfo {year} {2015})}\BibitemShut {NoStop}%
\bibitem [{\citenamefont {Mikhailov}(2015)}]{arXiv:1506.00534}%
  \BibitemOpen
  \bibfield  {author} {\bibinfo {author} {\bibfnamefont {S.~A.}\ \bibnamefont
  {Mikhailov}},\ }\href@noop {} {\enquote {\bibinfo {title} {Quantum theory of
  the third-order nonlinear electrodynamic effects of graphene},}\ } (\bibinfo
  {year} {2015}),\ \bibinfo {note} {arXiv:1506.00534},\ \Eprint
  {http://arxiv.org/abs/1506.00534} {1506.00534} \BibitemShut {NoStop}%
\bibitem [{\citenamefont {Zhou}\ \emph {et~al.}(2007)\citenamefont {Zhou},
  \citenamefont {Gweon}, \citenamefont {Fedorov}, \citenamefont {First},
  \citenamefont {de~Heer}, \citenamefont {Lee}, \citenamefont {Guinea},
  \citenamefont {Castro~Neto},\ and\ \citenamefont
  {Lanzara}}]{Nat.Mater._6_916_2007_Zhou}%
  \BibitemOpen
  \bibfield  {author} {\bibinfo {author} {\bibfnamefont {S.~Y.}\ \bibnamefont
  {Zhou}}, \bibinfo {author} {\bibfnamefont {G.-H.}\ \bibnamefont {Gweon}},
  \bibinfo {author} {\bibfnamefont {A.~V.}\ \bibnamefont {Fedorov}}, \bibinfo
  {author} {\bibfnamefont {P.~N.}\ \bibnamefont {First}}, \bibinfo {author}
  {\bibfnamefont {W.~A.}\ \bibnamefont {de~Heer}}, \bibinfo {author}
  {\bibfnamefont {D.-H.}\ \bibnamefont {Lee}}, \bibinfo {author} {\bibfnamefont
  {F.}~\bibnamefont {Guinea}}, \bibinfo {author} {\bibfnamefont {A.~H.}\
  \bibnamefont {Castro~Neto}}, \ and\ \bibinfo {author} {\bibfnamefont
  {A.}~\bibnamefont {Lanzara}},\ }\href {\doibase 10.1038/nmat2056} {\bibfield
  {journal} {\bibinfo  {journal} {Nat. Mater.}\ }\textbf {\bibinfo {volume}
  {6}},\ \bibinfo {pages} {916–916} (\bibinfo {year} {2007})}\BibitemShut
  {NoStop}%
\bibitem [{\citenamefont {Marzari}\ and\ \citenamefont
  {Vanderbilt}(1997)}]{Phys.Rev.B_56_12847_1997_Marzari}%
  \BibitemOpen
  \bibfield  {author} {\bibinfo {author} {\bibfnamefont {N.}~\bibnamefont
  {Marzari}}\ and\ \bibinfo {author} {\bibfnamefont {D.}~\bibnamefont
  {Vanderbilt}},\ }\href {\doibase 10.1103/physrevb.56.12847} {\bibfield
  {journal} {\bibinfo  {journal} {Phys. Rev. B}\ }\textbf {\bibinfo {volume}
  {56}},\ \bibinfo {pages} {12847} (\bibinfo {year} {1997})}\BibitemShut
  {NoStop}%
\bibitem [{\citenamefont {Boyd}(2008)}]{boyd_nonlinearoptics}%
  \BibitemOpen
  \bibfield  {author} {\bibinfo {author} {\bibfnamefont {R.~W.}\ \bibnamefont
  {Boyd}},\ }\href@noop {} {\emph {\bibinfo {title} {Nonlinear Optics}}},\
  \bibinfo {edition} {3rd}\ ed.\ (\bibinfo  {publisher} {Academic},\ \bibinfo
  {year} {2008})\BibitemShut {NoStop}%
\bibitem [{\citenamefont {Wu}\ \emph {et~al.}(2012)\citenamefont {Wu},
  \citenamefont {Mao}, \citenamefont {Jones}, \citenamefont {Yao},
  \citenamefont {Zhang},\ and\ \citenamefont {Xu}}]{NanoLett._12_2032_2012_Wu}%
  \BibitemOpen
  \bibfield  {author} {\bibinfo {author} {\bibfnamefont {S.}~\bibnamefont
  {Wu}}, \bibinfo {author} {\bibfnamefont {L.}~\bibnamefont {Mao}}, \bibinfo
  {author} {\bibfnamefont {A.~M.}\ \bibnamefont {Jones}}, \bibinfo {author}
  {\bibfnamefont {W.}~\bibnamefont {Yao}}, \bibinfo {author} {\bibfnamefont
  {C.}~\bibnamefont {Zhang}}, \ and\ \bibinfo {author} {\bibfnamefont
  {X.}~\bibnamefont {Xu}},\ }\href {\doibase 10.1021/nl300084j} {\bibfield
  {journal} {\bibinfo  {journal} {Nano Lett.}\ }\textbf {\bibinfo {volume}
  {12}},\ \bibinfo {pages} {2032} (\bibinfo {year} {2012})}\BibitemShut
  {NoStop}%
\bibitem [{\citenamefont {Guo}\ and\ \citenamefont
  {Lin}(2005)}]{Phys.Rev.B_72_075416_2005_Guo}%
  \BibitemOpen
  \bibfield  {author} {\bibinfo {author} {\bibfnamefont {G.~Y.}\ \bibnamefont
  {Guo}}\ and\ \bibinfo {author} {\bibfnamefont {J.~C.}\ \bibnamefont {Lin}},\
  }\href {\doibase 10.1103/PhysRevB.72.075416} {\bibfield  {journal} {\bibinfo
  {journal} {Phys. Rev. B}\ }\textbf {\bibinfo {volume} {72}},\ \bibinfo
  {pages} {075416} (\bibinfo {year} {2005})}\BibitemShut {NoStop}%
\bibitem [{\citenamefont {Gr\"uning}\ and\ \citenamefont
  {Attaccalite}(2014)}]{Phys.Rev.B_89_081102R_2014_Gruening}%
  \BibitemOpen
  \bibfield  {author} {\bibinfo {author} {\bibfnamefont {M.}~\bibnamefont
  {Gr\"uning}}\ and\ \bibinfo {author} {\bibfnamefont {C.}~\bibnamefont
  {Attaccalite}},\ }\href {\doibase 10.1103/physrevb.89.081102} {\bibfield
  {journal} {\bibinfo  {journal} {Phys. Rev. B}\ }\textbf {\bibinfo {volume}
  {89}},\ \bibinfo {pages} {081102(R)} (\bibinfo {year} {2014})}\BibitemShut
  {NoStop}%
\bibitem [{Note1()}]{Note1}%
  \BibitemOpen
  \bibinfo {note} {The situation may be different for calculations using
  $\protect \bm {p}\cdot \protect \bm {A}$ interaction. In our numerical
  calculation of linear response with $\protect \bm {p}\cdot \protect \bm {A}$
  interaction, we find that the inclusion of all $\protect \bm {k}$ in the
  whole Brillouin zone is necessary for the imaginary part of the linear
  conductivity, even though the inclusion of $\protect \bm {k}$ only with
  transition energy close to the photon energy is adequate for its real
  part.}\BibitemShut {Stop}%
\end{thebibliography}
%

\end{document}